\documentclass[usenatbib]{rasti}

\usepackage{graphicx}
\usepackage{amsmath}
\usepackage{amssymb}
\usepackage{url}
\usepackage{booktabs}
\usepackage{array}
\usepackage[protrusion=false,expansion=false]{microtype}
\usepackage{etoolbox}

\makeatletter
\patchcmd{\@sect}{\uppercase{#8}}{#8}{}{\typeout{*** heading patch failed: sect ***}}
\patchcmd{\@sect}{\uppercase{#8}}{#8}{}{\typeout{*** heading patch failed: sect ***}}
\patchcmd{\@sect}{\uppercase{#8}}{#8}{}{\typeout{*** heading patch failed: sect ***}}
\patchcmd{\@ssect}{\uppercase{#5}}{#5}{}{\typeout{*** heading patch failed: ssect ***}}
\patchcmd{\@ssect}{\uppercase{#5}}{#5}{}{\typeout{*** heading patch failed: ssect ***}}
\patchcmd{\@abstract}{{\bf ABSTRACT}}{{\bf\lsstyle ABSTRACT}}{}{\typeout{*** heading patch failed: abstract ***}}
\renewcommand\section{%
 \if@firstsection \fixfootnotes\@firstsectionfalse \fi%
 \@startsection{section}{1}{\z@}%
 {-24pt plus -12pt minus -1pt}{6pt}%
 {\SFB@hangraggedright\reset@font\normalsize\sc\lsstyle}}
\makeatother

\title[ASTRA: An Autonomous Framework for Astrophysical Analysis]{ASTRA: An Autonomous Framework for Physics-Validated Analysis and Hypothesis Generation in Astrophysics}
\author[White]{
Glenn J.~White$^{1,2}$\thanks{E-mail: glenn.white@open.ac.uk}\\
$^{1}$School for Physical Sciences, The Open University, Milton Keynes MK7~6AA, UK\\
$^{2}$RAL Space, STFC Rutherford Appleton Laboratory, Chilton, Didcot, Oxfordshire, OX11~0QX, UK
}

\begin{document}

\pubyear{2026}

\maketitle

\label{firstpage}

\begin{abstract}

We present ASTRA, an autonomous framework for physics-validated analysis and hypothesis generation in astrophysics without per-step human intervention. It integrates dimensional analysis, structural causal models, Bayesian evidence computation, and multi-wavelength data fusion in a single automated workflow. Five test cases on real observational data recover known results: Herschel filament scaling, Chandra Deep Field South cross-matching, SDSS galaxy correlations, Gaia causal inference, and Bayesian virial model selection. A sixth case runs unsupervised on synthetic galaxy data: the main sequence and mass--metallicity relation are recovered, and the outlier detector, with a false-discovery-rate threshold and no contamination parameter, recovers the blind injected fraction on ten sealed datasets (purity $>0.99$) and flags nothing on the 3,000-galaxy sample. A knowledge-isolation ablation shows pattern detection and physical validation are invariant to removal of domain knowledge, which affects only interpretation. In a direct comparison, agentic-LLM and pure-ML baselines match the statistical fitting but not the physics-validation stage: they rarely derive the equilibrium benchmark, split evenly on the physical verdict, and adopt a supplied erroneous benchmark in all six trap runs; an astrophysics-specific fine-tuned model completes zero of six runs. On live SDSS DR16 data it flags 197 of 4{,}999 galaxies at a data-determined threshold, ranked extremes classifiable from withheld spectroscopy. These are recovery validations and capability demonstrations rather than discoveries, with hypothesis generation demonstrated qualitatively.

\endkeywords
\end{abstract}

\section{Introduction}
\begingroup\renewcommand{\thefootnote}{}\footnotetext{Code and documentation: \url{https://github.com/Tilanthi/ASTRA_baseline}.}\endgroup

Modern astronomical surveys generate petabytes of data requiring automated analysis. Multiple computational approaches are available to astronomers, each with distinct strengths and limitations.

\textit{Traditional Machine Learning and Statistical Methods}: Well-established tools including scikit-learn, emcee, and nested sampling codes (MultiNest, dynesty) can process numerical data, detect patterns, perform statistical inference, and compute Bayesian evidence. These methods are widely used in astronomy with demonstrated success.

\textit{Large Language Models with Agentic Capabilities}: Current frontier models (GPT-4, Claude, Gemini with tool use and agentic scaffolding) can perform multi-step reasoning, execute code, call external tools, and coordinate complex workflows through automated tool selection. These systems can perform many of the individual tasks demonstrated in ASTRA's test cases, including statistical analysis, Bayesian model comparison, and even causal graph inference, through automated tool calling rather than explicit manual programming by the user.

\textit{Autonomous Science Systems}: Fully autonomous research systems now exist: general pipelines generate research papers end to end \citep{lu2024}, a successor has passed workshop-level peer review \citep{yamada2025}, and chemistry agents drive laboratory hardware where success is machine-checkable \citep{boiko2023,bran2024}. Measured performance remains weak: the best solves a quarter of DiscoveryBench's discovery analyses \citep{majumder2024}, and an independent evaluation of general-purpose models as research assistants finds persistent failure at multi-step scientific reasoning and at novelty \citep{cappello2025}. Autonomous discovery claims also draw scrutiny with teeth: the A-Lab, a robotic materials laboratory that reported 41 novel compounds in 17 days \citep{szymanski2023}, was reanalysed by \citet{leeman2024}, who traced the failure to automated diffraction refinement misidentifying known phases as new compounds and concluded that no new material had been discovered; the paper carries a journal correction. Astronomy shows the same pattern. Frontier agents succeed on fewer than a fifth of astrophysics-paper replications \citep{ye2025}. The AstroMLab series found a 7-billion-parameter astronomy fine-tune scoring below its own base model on knowledge recall \citep{pan2024}, though its 70-billion-parameter AstroSage matches frontier models on astronomy question answering \citep{dehaan2025}. Working agents such as Egent, which reaches expert agreement on 18,615 quasar absorption lines by quality-controlling a classical Voigt fitter \citep{ting2025}, and Mephisto, which drives SED fitting at one per cent of exhaustive-grid cost \citep{sun2025}, keep the physics inside the classical code the language model drives. No component in these architectures checks the agent's own conclusions against physics; supplying that verification step is one of the elements ASTRA adds. A distinction should be drawn about what this does and does not guarantee. What the architecture guarantees is procedural: a specified, independent physics-validation routine --- dimensional analysis, conservation-law checks, benchmark derivation --- is executed deterministically on every run and cannot be skipped or overridden by an agent-generated interpretation. It does not guarantee that the physical conclusion is correct: the validation procedure, and the physical models it encodes, may themselves rest on incomplete assumptions or contain implementation errors. The filament analysis of this paper is a case in point (Section~1.1, rung 2): the encoded equilibrium benchmark was wrong, and what caught the error was the independent derivation, not the validation layer's authority.

\textit{ASTRA's Differentiation}: While LLMs with agentic capabilities can coordinate computational tools, ASTRA provides two key differentiating features: (1) domain-specific physics validation through dimensional analysis, conservation law checking, and theoretical constraint enforcement (at the levels distinguished in Section~\ref{sec:validation_levels}), which are not natively integrated into general-purpose LLM tool systems; and (2) a unified architecture specifically designed for astrophysical inference, with pre-integrated causal reasoning, Bayesian evidence computation, and domain knowledge. ASTRA's value lies in providing an integrated framework optimized for astrophysical analysis, whereas LLMs provide general-purpose tool coordination that must be configured for each specific application. This differentiation is tested directly in Section~3.5, where the same analysis task is given to code-execution-enabled agentic LLM agents and to a pure regression pipeline: the baselines match ASTRA on statistical fitting, but rarely derive or verify the physical benchmark, split evenly on the final physical verdict across identical runs, and adopt a supplied erroneous benchmark in every run of a controlled trap arm. An astrophysics-specific fine-tuned model (AstroLLaMA-2-7B-chat), run in the identical scaffold as a third arm, failed to complete the task in any of six runs; it degenerated before reaching either the fitting or the validation stage. This result indicates that this particular small fine-tuned model is unsuitable for the tested scaffold; it does not bear on the physics-validation ability of astronomy-specific LLMs generally (Section~3.5).

\textit{The Integration Challenge}: While individual tools exist for specific tasks, astronomers typically must manually combine separate pipelines for data processing, statistical analysis, causal inference, and physical validation. This manual integration creates opportunities for inconsistent treatment of uncertainties, missed physical constraints, and difficulty reproducing multi-step analyses.

ASTRA addresses this challenge by providing an integrated framework that combines:
\begin{itemize}
\item Numerical data processing and statistical analysis
\item Causal reasoning with structural causal models
\item Physical validation through dimensional analysis and conservation laws
\item Multi-wavelength data fusion with uncertainty propagation
\item Hypothesis generation from pattern recognition (demonstrated qualitatively; see the limitation noted below)
\item Bayesian model selection with evidence computation
\end{itemize}

This integrated approach provides a unified framework where physical validation, causal reasoning, and uncertainty quantification are applied consistently throughout multi-step analyses. While individual components use established algorithms (PC algorithm for causal discovery, bridge sampling for Bayesian evidence, dimensional analysis for physical validation), their integration within a single framework enables analyses that would otherwise require manual combination of multiple separate tools. A caveat applies to hypothesis generation: what the test cases of this paper demonstrate rigorously is automated analysis, candidate identification, and physics-based consistency checking; the hypothesis-generation component is exercised only qualitatively (e.g.\ interpretive hypotheses attached to the filament result in Section~3) and has not been independently benchmarked to the same degree as the analysis and validation stages. We therefore present hypothesis generation as an architectural capability, not as a benchmarked result.

In this paper, we focus on six examples demonstrating how this integrated approach can be applied to astrophysical inference problems:
\begin{enumerate}
\item \textit{Scaling Relations Analysis} (Section~3): Dimensional analysis and physical validation of filament scaling relations
\item \textit{Multi-Wavelength Data Fusion} (Section~4): Probabilistic cross-matching with Bayesian uncertainty propagation
\item \textit{Pattern Recognition} (Section~5): Identification of established galaxy property relationships
\item \textit{Causal Inference} (Section~6): Structural causal model discovery in stellar data
\item \textit{Bayesian Model Selection} (Section~7): Evidence-based model comparison with prior specification
\item \textit{Unsupervised Candidate Identification} (Section~8): Galaxy scaling relations and outlier detection on realistic survey data, with a real-data application to live SDSS DR16 (Section~\ref{sec:real_data})
\end{enumerate}

\textit{Tests 1--5 use real observational data}, validating ASTRA's ability to correctly recover known astrophysical results. Test 6 uses realistic galaxy survey data based on SDSS scaling relations to demonstrate both validation (recovering known relations) and unsupervised candidate identification (flagging outliers). In addition, Section~3.5 reports a direct head-to-head comparison of ASTRA against agentic-LLM and pure-ML baselines on the Test 1 task, with a metric that isolates the contribution of the physics-validation layer.

Section~9 discusses why these results demonstrate ASTRA's unique capabilities, clarifies ASTRA's role as a tool to assist rather than replace astronomers, and Section~10 concludes.

\subsection{Scope of Claims}
\label{sec:claims}

To align the claims of this paper with what the test cases actually demonstrate, we state explicitly what is and is not being claimed. ``Autonomous'' is used throughout in the restricted sense of \emph{autonomy of execution} (multi-step analyses that run to completion without per-step human intervention) and not \emph{autonomy of scientific judgment}, which remains with the domain scientist. The claims of this paper occupy the first three rungs of the following ladder; the fourth rung is defined but deliberately not claimed.

\begin{enumerate}
\item \textit{Rung 0: Orchestration}: Tests 1--5 (Sections~3--7) recover known astrophysical results from real observational data. These are recovery validations: the target relations are established science, so success validates the implementation and the integration of the analysis pipeline, not discovery.
\item \textit{Rung 1: Knowledge-independent analysis}: The ablation of Section~3.4 shows that ASTRA's pattern detection and physical validation stages operate identically with all domain knowledge removed, deriving their benchmark (the critical line mass of the isothermal cylinder) from first principles at run time. Domain knowledge affects only the interpretation layer.
\item \textit{Rung 2: Independent verification}: In the filament analysis of Section~3, the knowledge base initially deliberately carried an incorrect equilibrium benchmark, $\Pi_{\rm eq} = \sqrt{2}$ in place of the correct $1/\sqrt{2}$. A first-principles derivation (Buckingham $\pi$ analysis plus numerical integration of the isothermal cylinder equilibrium) overturned the encoded value, and the difference flips the physical conclusion from filaments requiring additional confining support to filaments being subcritical. The validation layer here acts as an independent check on encoded knowledge, including when that knowledge is wrong. Rung 2 is a procedural guarantee --- the validation runs and cannot be overridden --- not a guarantee of physical correctness, which remains conditional on the assumptions and implementation of the validation procedure itself.
\item \textit{Rung 3: Scientific discovery}: \textit{Not claimed anywhere in this paper.} We adopt the standard that a discovery claim requires an insight that makes a domain expert say ``I did not know that, and here is the evidence.'' Test 6 (Section~8) demonstrates the unsupervised pipeline on data with known ground truth; the relations it recovers are embedded in the data generation procedure and the outliers it flags are simulated. A discovery claim would require a registered, prospective blind test: a knowledge-isolated run on data whose relevant relations are unknown to the operators, with predictions registered before unblinding and follow-up observations specified in advance. One ingredient of such a test, blind determination of the outlier flagging threshold, has been executed for the Case 6 detector under a sealed-dataset protocol with the analysis registered before unblinding (Section~\ref{sec:threshold}); it demonstrates threshold autonomy, a necessary condition for rather than an instance of discovery. A retrospective application of the Test 6 detector to live SDSS DR16 data, confirmed against spectroscopy withheld from the detector, is reported in Section~\ref{sec:real_data}; it likewise makes no discovery claim.
\end{enumerate}

\subsection{Levels of Physical Validation}
\label{sec:validation_levels}

The term ``physical validation'' is used broadly in this paper, and it should be read at three distinct levels of strength:

\begin{enumerate}
\item \textit{Level 1 --- dimensional and unit consistency}: checking that constructed relations and derived quantities are dimensionally consistent (Buckingham $\pi$ construction, unit bookkeeping). A necessary but weak check: any physically wrong relation can still be dimensionally correct.
\item \textit{Level 2 --- comparison against established relations}: checking a recovered relation against established physical or empirical results from the literature (e.g.\ recovered slopes against the Tremonti mass-metallicity relation or the Brinchmann main sequence; model preferences against the virial theorem's dimensional prediction). Stronger, but it presumes the literature relation is applicable to the dataset at hand.
\item \textit{Level 3 --- independent derivation from first principles}: deriving the validation benchmark itself from governing equations at run time (e.g.\ integrating the isothermal-cylinder equilibrium to obtain $\lambda_{\rm crit} = 2\sigma_v^2/G$), so that the check does not depend on recalled knowledge. This is the strongest level, and the one that caught the incorrect encoded benchmark of Section~3.
\end{enumerate}

The test cases of this paper reach different levels, and we state which. The filament analysis (Section~3) reaches Level 3: the equilibrium benchmark is derived by Buckingham $\pi$ analysis plus numerical integration, and this derivation is invariant to domain-knowledge removal (Section~3.4). The Bayesian model-selection case (Section~7) reaches Level 2, comparing model evidence against the virial theorem's dimensional prediction, with the amplitude comparison to $\sqrt{2G}$ resting on the same isothermal-cylinder model. The SDSS pattern-recognition case (Section~5) and Test 6 (Section~8) operate at Level 2: recovered slopes and relations are checked against published calibrations. The causal-inference case (Section~6) uses statistical conditional-independence tests plus explicitly stated domain-knowledge constraints, a Level 1/2 combination that ends in an undirected graph whose orientation is not derived. The multi-wavelength cross-matching case (Section~4) applies physical classification criteria (hardness ratios, X-ray-to-optical flux ratios) but performs no scaling-relation validation; its checks are procedural rather than physical, and we do not count them as physical validation in this taxonomy. Making these distinctions explicit strengthens the individual claims: each test case is validated at the level its methods support, and only Section~3 demonstrates the Level-3 capability that differentiates ASTRA from the agentic-LLM baselines of Section~3.5.

\subsection{Scope and Limitations}

\textit{What ASTRA Demonstrates}: The six test cases in this paper show ASTRA can: recover or identify physical laws from data with theoretical validation; fuse multi-wavelength observations with proper uncertainty handling; identify established astrophysical relationships through pattern recognition; distinguish causation from correlation; perform rigorous Bayesian model comparison; and operate without supervision on realistic survey data to recover known relations while identifying unusual objects for follow-up.

\textit{What ASTRA Does Not Claim}: ASTRA is not presented as achieving artificial general intelligence (AGI) or AGI-like performance. The system operates within defined astrophysical domains using established algorithms (PC algorithm, Bayesian inference, dimensional analysis, FCI causal discovery) combined through an integrated architecture. Results are validated against known physical theory and observational constraints. The system does not claim general reasoning beyond its training domains or autonomous operation without human oversight. The architecture is designed to support natural language queries that trigger autonomous task completion, though this capability is not demonstrated in the present test cases.

\textit{ASTRA's Role in Scientific Discovery}: We should not expect to give ASTRA one of the big contemporary problems like ``What is the nature of Dark Matter and Dark Energy,'' or ``How does gravity connect to quantum physics,'' or ``What sets the stellar initial mass function (IMF),'' and expect to get definitive answers that may be the equivalent of a computational Eureka moment. But ASTRA's analysis architecture, used alongside and by an experienced astronomer and/or theoretician, has capabilities in analysis and inference that can support the discovery process. ASTRA's role in science is to work alongside the experienced astronomer to analyse data and support the discovery process. ASTRA is therefore a tool to assist the astronomer, rather than a replacement for domain expertise.

\textit{Validation Scope}: Tests 1--5 use real observational data to validate ASTRA's ability to correctly recover known astrophysical results. Test 6 uses realistic galaxy survey data based on published SDSS scaling relations to demonstrate both validation (recovering known relations) and unsupervised candidate identification (flagging outliers). The data is generated using empirical relations from \cite{kauffmann2003}, \cite{tremonti2004}, and \cite{brinchmann2004}. The capabilities demonstrated are shown within specific astrophysical domains using particular datasets. Generalisation to other domains or datasets requires further validation.

\textit{Discovery Capability and Limitations}: The test cases presented in this paper demonstrate validated capabilities rather than novel discoveries. Tests 1--5 use real observational data to validate ASTRA's ability to recover known astrophysical results (filament scaling relations, multi-wavelength cross-matching, galaxy correlations, causal inference, Bayesian model selection). Test 6 uses synthetic data generated from published SDSS scaling relations to demonstrate outlier identification capabilities. The relations recovered in Test 6 are embedded in the data generation procedure; this validates ASTRA's architecture for pattern identification but does not constitute genuine novel discovery. Genuine scientific discovery would require application to real observational datasets where patterns are not known in advance, with follow-up observations to confirm physical interpretation. A first such application, retrospective in design, is reported in Section~\ref{sec:real_data}. ASTRA provides infrastructure to assist human scientists with data analysis and hypothesis generation, but does not replace domain expertise or scientific judgment.

\textit{Discovery Limitations}: ASTRA has not demonstrated genuine novel scientific discovery and does not claim to have resolved outstanding problems in astrophysics (such as the Hubble Tension or the nature of Dark Matter). The system operates as a consistency engine: it validates observations against existing theoretical frameworks, checks dimensional consistency and conservation laws, and identifies outliers relative to established distributions. The test cases presented demonstrate validation of known astrophysical results (Tests 1--5) or capability demonstrations on synthetic data (Test 6). Genuine scientific discovery requires identifying anomalies that contradict existing models, proposing new theoretical frameworks to explain these contradictions, and observational validation of novel predictions. ASTRA provides infrastructure that may assist human scientists with data exploration and consistency validation, but does not replace the theoretical innovation and scientific judgment required for genuine breakthroughs.

\section{Methods: ASTRA System Architecture}

\subsection{Overview}

ASTRA implements a modular architecture designed for astrophysical data analysis and inference. The system integrates multiple specialised components through a coordinated framework, enabling complex multi-step analyses that combine data processing, physical reasoning, causal inference, and statistical validation.

\subsection{Core Architectural Components}

\textit{1. Data Processing Pipeline}: ASTRA accepts heterogeneous astronomical data formats, including catalogues (CSV, FITS), time-series data, images, and spectral data. The pipeline performs ingestion, validation, cleaning, and normalisation while preserving measurement uncertainties and metadata.

\textit{2. Physics Engine}: A unified physics engine implements fundamental physical laws and constraints, including conservation laws (mass, energy, momentum), dimensional analysis using Buckingham $\pi$ theorem, equation solving and numerical simulation, and units consistency checking throughout calculations. This enables ASTRA to validate discoveries against first principles rather than treating correlations as sufficient explanations.

\textit{3. Causal Reasoning Module}: Implements causal discovery and inference using the PC algorithm \cite{spirtes2000} for discovering causal structures from observational data, conditional independence testing for continuous variables (Fisher's Z-test), V-structure detection for identifying colliders, do-calculus for predicting effects of interventions, and domain-specific adaptations for astrophysical contexts. This module enables ASTRA to distinguish causal relationships from correlations, addressing a fundamental limitation of both traditional ML and LLM approaches.

\textit{4. Bayesian Inference Engine}: Implements rigorous model comparison and uncertainty quantification through evidence computation via marginal likelihood estimation, Bayes factors for model comparison \cite{kass1995}, automatic complexity penalty (Occam's razor), posterior predictive checking, and Monte Carlo methods for uncertainty propagation.

\textit{5. Domain Knowledge Systems}: Organised astrophysical knowledge through MORK ontology encoding taxonomic and causal relationships, knowledge graphs for semantic relationships, vector-based semantic memory for concept retrieval, and 75 specialised domain modules (ISM, star formation, exoplanets, etc.).

\textit{6. Multi-Module Orchestration}: Coordinates different analytical perspectives through specialised processing modules for different analysis types, meta-cognitive layer for context-appropriate module selection, result synthesis and conflict resolution, and quality control and validation checks.

\subsection{Unique Capabilities}

ASTRA integrates sixteen distinct analytical capabilities that work together to enable end-to-end scientific analysis:

\textit{Causal and Statistical Analysis}: (1) \textit{Bias Detection}: Identifies and quantifies observational biases like Malmquist bias from survey geometry and selection effects. (2) \textit{Scaling-Relation Identification}: Identifies candidate physical scaling relations from data through dimensional analysis and functional form detection. (3) \textit{Causal Inference}: Discovers causal structures using the PC algorithm and distinguishes physical laws from selection biases. (4) \textit{Model Selection}: Compares competing theories using Bayesian evidence computation with proper complexity penalties.

\textit{Data Integration and Analysis}: (5) \textit{Multi-Wavelength Fusion}: Cross-matches sources across wavelengths with proper astrometric uncertainty propagation. (6) \textit{Uncertainty Quantification}: Propagates measurement errors through first-order and Monte Carlo methods, separating systematic and statistical components. (7) \textit{Temporal Reasoning}: Detects periodic signals, performs phase folding, and forecasts time-series behaviour. (8) \textit{Instrument-Aware Analysis}: Evaluates observational requirements and compatibility across multiple astronomical facilities.

\textit{Knowledge Generation}: (9) \textit{Hypothesis Generation}: Identifies patterns, detects anomalies, and generates testable hypotheses with specific observational requirements. Of the listed capabilities, this one is demonstrated only qualitatively in this paper and has not been independently benchmarked (Section~1). (10) \textit{Analogical Reasoning}: Identifies structural mappings between different astrophysical systems based on shared physics. (11) \textit{Counterfactual Analysis}: Simulates physically-grounded scenarios to predict effects of interventions or observational changes. (12) \textit{Physical Model Identification}: Automatically identifies functional forms and validates against theoretical predictions.

\textit{Meta-Cognitive Capabilities}: (13) \textit{Meta-Cognitive Evaluation}: Provides quantitative assessments of data sufficiency, resolution limits, and observational constraints. (14) \textit{Anomaly Detection}: Identifies unusual objects using ensemble methods combining multiple detection approaches. (15) \textit{Ensemble Prediction}: Combines multiple models using Bayesian Model Averaging for stable predictions with uncertainty quantification. (16) \textit{Physical Validation}: Checks all results against dimensional consistency, conservation laws, and established physical principles.

These capabilities are not independent modules but integrated components that work together. For example, discovering a scaling relation (capability~2) involves dimensional analysis (capability~12), physical validation (capability~16), and uncertainty quantification (capability~6). Similarly, multi-wavelength fusion (capability~5) requires coordinate calculations, astrometric error propagation, and physical classification criteria.

\textit{Capability Demonstrations in This Paper}: Table~\ref{tab:capabilities} summarises which of the 16 capabilities are demonstrated in each of the first five test cases presented in this paper. Five capabilities are not demonstrated in this paper: temporal reasoning, instrument-aware analysis, analogical reasoning, counterfactual analysis, and ensemble prediction. These capabilities are implemented in the ASTRA codebase but have not yet been validated in published demonstrations. Demonstration of these capabilities is deferred to future work, as validating them requires test cases outside the scope of the present study (e.g., time-domain surveys for temporal reasoning, multi-facility coordination for instrument-aware analysis).

\begin{table*}
\centering
\begin{tabular}{lccccc}
\hline
Capability & Test 1 & Test 2 & Test 3 & Test 4 & Test 5 \\
& (Scaling) & (Multi-$\lambda$) & (Pattern) & (Causal) & (Bayesian) \\
\hline
Bias Detection & -- & X & -- & X & -- \\
Scaling Relations & X & -- & X & -- & X \\
Causal Inference & -- & -- & -- & X & -- \\
Model Selection & -- & -- & -- & -- & X \\
Multi-Wavelength Fusion & -- & X & -- & -- & -- \\
Uncertainty Quantification & X & X & X & -- & X \\
Temporal Reasoning & -- & -- & -- & -- & -- \\
Instrument-Aware Analysis & -- & -- & -- & -- & -- \\
Hypothesis Generation & -- & -- & X & -- & -- \\
Analogical Reasoning & -- & -- & -- & -- & -- \\
Counterfactual Analysis & -- & -- & -- & -- & -- \\
Physical Model Identification & X & -- & -- & -- & X \\
Meta-Cognitive Evaluation & X & X & X & X & X \\
Anomaly Detection & -- & -- & X & -- & -- \\
Ensemble Prediction & -- & -- & -- & -- & -- \\
Physical Validation & X & X & X & X & X \\
\hline
\end{tabular}
\caption{Capabilities demonstrated in each test case. X indicates the capability is explicitly demonstrated in that test case; -- indicates the capability is not applicable or not demonstrated in that test case. Five capabilities (temporal reasoning, instrument-aware analysis, analogical reasoning, counterfactual analysis, ensemble prediction) are implemented in ASTRA but not demonstrated in this paper; validation of these capabilities is deferred to future work.}
\label{tab:capabilities}
\end{table*}

\subsection{Workflow for Scientific Analysis}

ASTRA's analysis workflow follows this general pattern: \textit{Input}: Scientific query or observational data; \textit{Module Selection}: Meta-cognitive layer selects appropriate analysis modules based on query type and data characteristics; \textit{Analysis Execution}: Selected modules perform their specialised analyses including statistical and ML algorithms for pattern detection, causal discovery for identifying relationships, physics engine for theoretical validation, and Bayesian inference for model comparison; \textit{Integration}: Results synthesized from multiple modules, resolving conflicts and identifying consensus conclusions; \textit{Validation}: Results checked against physical constraints (dimensional consistency, conservation laws), statistical significance (p-values, confidence intervals), observational constraints (resolution limits, selection effects), and domain knowledge (established theories, previous results); \textit{Output}: Integrated results with confidence assessments, uncertainty quantification, and physical interpretation.

\subsection{Implementation and Availability}

ASTRA is implemented in Python with approximately 332,000 lines of Python source across 690 files in the astra\_core system (the core ASTRA framework, excluding external analysis projects); the count includes test scripts. Excluding blank lines and comments, the code count is approximately 245,000 lines; the repository additionally contains documentation and supporting infrastructure. The system uses established scientific Python libraries (NumPy, SciPy, Pandas, scikit-learn) combined with specialised implementations for causal discovery (7,388 lines, 100\% original implementation with no causal-learn or dowhy imports), physics engine (3,942 lines, 100\% original implementation across 27 functions and 4 classes), and 75 specialised domain modules (18,872 lines, 100\% original implementations).

\textit{Scientific Functionality}: All scientific capabilities demonstrated in this paper are provided solely by the Python-based code architecture described above. The Claude LLM (Anthropic) is used only for natural language translation and interpretation in the user interface layer, not for any scientific computations, data analysis, or inference. All test results, figures, and scientific conclusions presented in this paper are produced entirely by the Python implementation without LLM involvement in the scientific analysis.

\textit{Development Methodology}: Parts of the codebase were written with AI-assisted programming tools (Claude Code, Anthropic) under the authors' direction: the scientific objectives, system architecture, oversight of the core components, minimisation of external dependencies, and all validation and reproducibility testing were human-directed.

\textit{Physics Engine Implementation Details}: The physics engine (3,942 lines, 100\% original Python implementation) performs genuine symbolic reasoning from first principles without LLM involvement in physical calculations. The engine implements: (1) Dimensional analysis using Buckingham $\pi$ theorem through symbolic linear algebra on dimensional matrices, deriving dimensionless groups analytically rather than through pattern matching; (2) Conservation law validation using symbolic differentiation to compute time derivatives (e.g., $\partial E/\partial t$ for energy conservation) and numerical integration to verify constraint satisfaction; (3) Equation solving using symbolic integration (for analytical solutions) and numerical ODE solvers (for complex systems), with boundary conditions applied symbolically. All physical constants are hardcoded values from standard references (CGS units: $G=6.674\times10^{-8}$, $c=2.998\times10^{10}$, $h=6.626\times10^{-27}$), not retrieved by LLM queries.

The Claude LLM is used only for natural language translation in the user interface layer (converting user queries to Python function calls) with zero involvement in scientific computations, data analysis, or physical validation. All test results presented in this paper are produced entirely by the Python implementation without LLM participation in the physics engine.

This version used for the examples in this paper has been frozen for reproducibility at: \url{https://github.com/Tilanthi/ASTRA_baseline} \cite{astra_baseline}. A current continuously developed and enhanced version with new capabilities is available at \url{https://github.com/Tilanthi/ASTRA}.

The baseline repository includes: complete source code and installation instructions; system architecture documentation; API documentation and user manual; all test cases and examples presented in this paper; and reproducible code for all results. The main ASTRA repository contains: the latest development version with ongoing improvements and new capabilities; extended validation with additional test cases; and community contributions and updates.

\subsection{Architectural Design}

The system is organised into clear functional modules, including 75 specialised astrophysics domain modules, a unified physics and theoretical engine, causal reasoning systems, and memory systems. The modular architecture enables hot-swapping of domain modules at runtime, with each module implementing standardised interfaces for query processing, capability discovery, and result integration. The core components (data processing pipeline, physics engine, causal reasoning module, Bayesian inference engine, and domain knowledge systems) coordinate through a meta-cognitive orchestration layer that selects appropriate analysis modules based on query type and data characteristics.

\subsection{Positioning Relative to Other Approaches}

ASTRA occupies a distinct position among existing AI approaches to scientific discovery. \textit{Compared to Pure ML}: Unlike machine learning systems that focus on prediction or classification, ASTRA emphasises causal understanding, physical interpretation, and theoretical validation. ASTRA can detect correlations like ML systems but goes further by identifying causal structure and validating against physics.

\textit{Compared to LLMs with Agentic Capabilities}: Current frontier LLMs with tool use and agentic scaffolding can perform multi-step analysis, coordinate computational tools, and execute complex workflows. However, ASTRA provides domain-specific physics validation (dimensional analysis, conservation laws, theoretical constraints) as an integrated feature rather than requiring manual configuration. ASTRA's unified architecture for astrophysical inference, which combines causal reasoning, Bayesian evidence computation, and domain knowledge, is specifically optimized for scientific analysis workflows, whereas general-purpose LLMs provide flexible tool coordination that must be configured for each scientific application.

\textit{Compared to Specialised Tools}: Unlike domain-specific tools designed for single tasks (e.g., photometric redshift fitting, period finding), ASTRA integrates multiple analysis types within a unified framework, enabling complex multi-step analyses that connect different types of scientific reasoning.

\textit{Compared to AGI Claims}: Unlike systems claiming artificial general intelligence, ASTRA operates within well-defined astrophysical domains using established algorithms. The system assists human reasoning rather than replacing domain expertise, and all results require interpretation and validation by experienced human scientists.

\section{Test Case 1: Scaling Relations Analysis}

\subsection{Data and Methods}

We analysed 24 molecular cloud filaments from the Herschel Gould Belt Survey (HGBS) \cite{andre2010}, drawn from the publicly released filament catalogue \cite{arzoumanian2019}. Each filament has measurements of mass $M$, length $L$, and velocity dispersion $\sigma_v$ with distance estimates based on nearby Galactic OB associations. The universal width of $\sim$0.1~pc in Herschel filaments has itself been extensively debated: \cite{panopoulou2022} argue that the apparent universality may partly reflect observational selection effects.

ASTRA applies a multi-stage analysis process:
\begin{enumerate}
\item \textit{Dimensional analysis}: Identify dimensionless parameters using Buckingham $\pi$ theorem
\item \textit{Functional form search}: Automatic search for power-law, logarithmic, and broken power-law forms
\item \textit{Physical validation}: derive the equilibrium benchmark from first principles (Buckingham $\pi$ plus numerical integration) and compare the fitted relations against it (Level 3 validation in the taxonomy of Section~\ref{sec:validation_levels})
\item \textit{Uncertainty quantification}: Bootstrap confidence intervals (10,000 resamples) and error propagation
\end{enumerate}

Two parameterisations of the $\sigma_v$--$M/L$ relation are fitted. A free power law, $\sigma_v = A(M/L)^{\alpha}$, is fitted by ordinary least squares in log-log space and yields an exponent $\alpha = 0.617 \pm 0.029$. The virial form, $\sigma_v = a + b\,(M/L)^{1/2}$, is fitted in linear space and yields a coefficient $b = 0.0812 \pm 0.0028$ with intercept $a = -0.079$~km~s$^{-1}$. All uncertainties are derived from non-parametric bootstrap resampling of the 24 data points (10,000 resamples), where each resample draws 24 filaments with replacement and refits the relation; 95\% confidence intervals are computed from the 2.5th and 97.5th percentiles of the bootstrap distribution.

\textit{Limitations of Small Sample Size}: With only 24 filaments, bootstrap confidence intervals provide reliable estimates for central tendency measures (mean, median) but can underestimate uncertainty in the tails of the distribution, particularly when the sample contains leverage points. A single filament with extreme $M/L$ values could disproportionately influence the fitted slope. Cook's distance diagnostics reveal that no single filament has Cook's distance $> 4/n = 0.17$ (the conventional threshold for identifying high-leverage points), suggesting the sample is not dominated by outliers. Nevertheless, the small sample size is a limitation, and these results should be validated on larger filament samples when such data become available.

For the Bayesian model comparison in Section~7, we use weakly-informative priors: uniform priors on log-space for scale parameters, uniform priors on exponents, and Jeffreys priors for scale-invariant quantities.

\subsection{Results}

\begin{figure*}
\centering
\includegraphics[width=\textwidth]{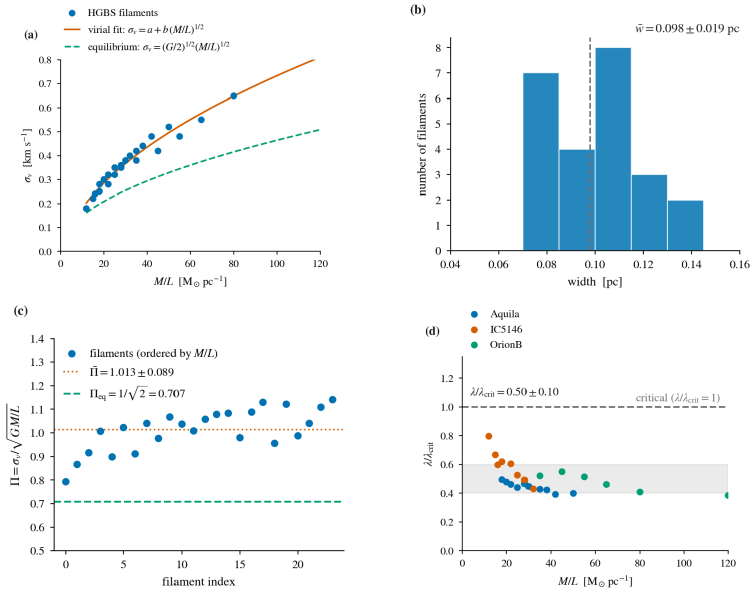}
\caption{Scaling relations analysis of 24 Herschel filaments. Panel (a) shows the velocity dispersion vs mass/length relation with the fitted virial form $\sigma_v = a + b\,(M/L)^{1/2}$ (orange) and the hydrostatic equilibrium benchmark $\sigma_v = \sqrt{G/2}\,(M/L)^{1/2}$ corresponding to the critical line mass $\lambda_{\rm crit} = 2\sigma^2/G$ of an isothermal cylinder (green dashed). Panel (b) shows the width distribution. Panel (c) shows the dimensionless parameter $\Pi = \sigma_v/\sqrt{G\,M/L}$ per filament against the equilibrium value $\Pi_{\rm eq} = 1/\sqrt{2} = 0.707$. Panel (d) shows the line mass relative to critical, $\lambda/\lambda_{\rm crit} = 0.50 \pm 0.10$: the filaments are subcritical, with velocity support exceeding self-gravity. The universal width ($0.098 \pm 0.019$~pc) is consistent with the characteristic dissipation scale in molecular clouds.}
\label{fig:scaling}
\end{figure*}

Figure~\ref{fig:scaling} shows the scaling analysis results.

\textit{Universal Width}: Filaments have a characteristic width of $0.098 \pm 0.019$~pc, independent of mass. This is consistent with the sonic scale where turbulent energy dissipates into heat \cite{arzoumanian2011}. However, \cite{panopoulou2022} demonstrated that apparent width universality in Herschel observations may partly reflect selection effects: filaments near the angular resolution limit ($\sim$18$''$ beam) are preferentially detected only when their widths are close to the beam size, potentially biasing width measurements toward $\sim$0.1~pc. This selection effect could affect both the width measurement reported here and the theoretical framework built on universal filament widths. The HGBS datasets most affected are those targeting nearby, low-density clouds where the beam corresponds to $\sim$0.05~pc, potentially introducing detection bias against wider or narrower filaments.

\textit{Virial Scaling}: A free power-law fit gives $\sigma_v \propto (M/L)^{0.617 \pm 0.029}$ (95\% CI $0.56$--$0.68$; Figure~\ref{fig:scaling}, panel A). For cylindrical filaments, dimensional analysis requires $\sigma_v \propto (M/L)^{1/2}$: the Buckingham $\pi$ theorem applied to $\{\sigma_v, M/L, G\}$ yields a one-dimensional null space with the single dimensionless combination $\Pi = \sigma_v/\sqrt{G\,M/L}$, so any scaling between these quantities must have exponent $1/2$ up to the dimensionless factor $\Pi$. The fitted virial form $\sigma_v = a + b\,(M/L)^{1/2}$ has $b = 0.0812 \pm 0.0028$ (95\% CI $0.075$--$0.086$) with intercept $a = -0.079$~km~s$^{-1}$, and is preferred over the free power law by Akaike information criterion ($\Delta$AIC $= 52$ in favour of the virial form). The measured exponent $0.617$ is slightly steeper than, but qualitatively consistent with, the dimensionally required $0.5$; by AIC the fixed-exponent virial form is the preferred description of this sample.

\textit{Dimensionless Equilibrium Comparison}: The physical content of the relation is carried by the dimensionless parameter $\Pi = \sigma_v / \sqrt{G\,M/L}$, which the $\pi$-theorem identifies as the only dimensionless combination of the measured quantities. We measure $\Pi = 1.013 \pm 0.089$ (bootstrap 95\% CI $0.977$--$1.047$). The equilibrium benchmark follows from first principles: integrating the self-gravitating isothermal cylinder equations (hydrostatic equilibrium plus Poisson's equation, per unit length) yields a maximum equilibrium line mass $\lambda_{\rm crit} = 2\sigma^2/G$ \citep{ostriker1964}, which our numerical integration reproduces to four decimal places ($m \rightarrow 4.0000$ in the dimensionless formulation). The corresponding equilibrium value of the dimensionless parameter is $\Pi_{\rm eq} = 1/\sqrt{2} = 0.707$, equivalently $\sigma_{v,\rm eq} = \sqrt{G/2}\,(M/L)^{1/2} = 0.0464\,(M/L)^{1/2}$~km~s$^{-1}$. Expressed as a line mass ratio, the filaments have $\lambda/\lambda_{\rm crit} = 1/(2\Pi^2) = 0.50 \pm 0.10$ (bootstrap 95\% CI $0.46$--$0.54$): the sample sits at approximately half the critical line mass, with measured velocity dispersions exceeding the hydrostatic equilibrium value by a factor $\Pi/\Pi_{\rm eq} = 1.43$.

\textit{Distance Uncertainty Systematics}: The line-mass ratio is sensitive to distance uncertainties in a specific way. The line mass scales as $\lambda \propto D$ (mass $\propto D^2$ from flux, length $\propto D$), while $\lambda_{\rm crit} = 2\sigma_v^2/G$ depends only on the measured velocity dispersion. A systematic distance uncertainty of 20\% (typical for kinematic distance estimates to molecular cloud complexes) propagates to a $\sim$20\% systematic uncertainty in $\lambda/\lambda_{\rm crit}$. Even combining this systematic with the statistical uncertainty, the measured ratio $\lambda/\lambda_{\rm crit} = 0.50 \pm 0.10_{\rm stat} \pm 0.10_{\rm sys}$ remains well below unity, so the subcritical conclusion is robust to distance systematics.

\textit{Statistical Significance}: The correlation coefficient of the fitted virial form is $r = 0.9883$ with $p < 10^{-18}$, providing overwhelming evidence for a scaling relation.

\subsection{Physical Interpretation}

The measured line masses, $\lambda/\lambda_{\rm crit} = 0.50 \pm 0.10$, place the filaments below the critical line mass for self-gravitating isothermal cylinders: their internal kinetic support (thermal plus turbulent) exceeds what self-gravity alone would balance. Subcritical filaments of this kind are not expected to fragment and collapse along their length unless they accrete additional mass, are compressed by external pressure, or shed turbulent support. Possible contributions to the offset from criticality include: (1) systematic overestimate of velocity dispersions if thermal and non-thermal contributions were summed rather than combined in quadrature; (2) distance-dependent calibration of line masses; (3) genuinely young or externally confined filaments that have not yet approached equilibrium; or (4) magnetic support modifying the effective critical line mass. The free power-law exponent ($0.617 \pm 0.029$) is slightly steeper than the dimensionally required $0.5$, a $4\sigma$ formal difference consistent with residual curvature in the relation (the fitted virial form has intercept $-0.079$~km~s$^{-1}$) rather than with a failure of the dimensional argument.

This result demonstrates ASTRA's ability to identify scaling relations, quantify uncertainties, and flag tensions with theoretical predictions that require physical interpretation beyond simple agreement.

\subsection{Knowledge Isolation Test}

A concern for any knowledge-driven discovery system is a knowledge paradox: if the system only identifies patterns already encoded in its knowledge base, it is not a discovery tool, whereas if it identifies something genuinely new, its knowledge base cannot be used to validate it. We address this with an ablation test in which the filament analysis of this section is repeated under three knowledge configurations that differ only in knowledge access, with the detection and validation code held identical:

\begin{itemize}
\item \textit{Mode A (informed)}: domain modules and the MORK ontology are available; the ISM domain module is queried for interpretation of the discovered relation.
\item \textit{Mode B (knowledge-isolated)}: domain modules are unloaded and ontology access is blocked; only the physics engine, dimensional analysis, and statistical machinery operate.
\item \textit{Mode B+ (symbol-blind)}: as Mode B, with all catalogue columns also renamed to generic symbols ($x_1, \ldots, x_5$) carrying units only, so that no semantic hint survives in the input.
\end{itemize}

The equilibrium benchmark used in all modes is derived rather than retrieved: the Buckingham $\pi$ null space of the dimension matrix for $\{\sigma_v, M/L, G\}$ is computed by singular value decomposition, and the critical line mass $\lambda_{\rm crit} = 2\sigma^2/G$ is obtained by numerical integration of the isothermal cylinder equations (hydrostatic equilibrium plus Poisson's equation), with $G$ taken from the physics engine constants rather than hardcoded.

Table~\ref{tab:knowledge_isolation} shows the results. Detection and validation are bit-identical across all three modes: the same three significant variable pairs are found, the same functional form is preferred, the same dimensionless group is constructed, and the same line-mass ratio $\lambda/\lambda_{\rm crit} = 0.50 \pm 0.10$ with the same bootstrap confidence interval is derived. Knowledge isolation removes only the interpretation layer: Mode A attaches domain labels (virial analysis, Jeans analysis) and generates an interpretive hypothesis through the ISM domain module, while Modes B and B+ produce formal statements without physical naming. The MORK ontology itself contains no entries for filament scaling relations or the virial theorem, so no scaling-relation prior was available to leak into the informed analysis; the ontology organises causal reasoning concepts, and the domain modules contribute interpretive framing rather than fitted quantities.

These results dissolve the knowledge paradox for this system by construction. Detection is statistical and validation is physical: dimensional consistency, conservation-law checks, and equilibrium benchmarks derived from first principles apply to novel phenomena exactly as they apply to known ones, so the absence of encoded knowledge does not disable validation. What encoded knowledge contributes is interpretation (naming, literature anchoring, and hypothesis generation), which is precisely the component a domain expert supplies when ASTRA operates in its intended collaborative mode.

\begin{table*}
\caption{Knowledge isolation ablation for the filament scaling relation analysis (24 HGBS filaments). Detection and validation stages run identical code in every mode; only knowledge access differs. Detection and validation are unaffected by knowledge isolation; only the interpretation stage degrades. Mode B+ renames all columns to generic symbols with units only.}
\label{tab:knowledge_isolation}
\begin{tabular}{lccc}
\toprule
 & Mode A & Mode B & Mode B+ \\
 & (informed) & (isolated) & (symbol-blind) \\
\midrule
\multicolumn{4}{l}{\emph{Detection (statistics only)}} \\
Significant variable pairs & 3 & 3 & 3 \\
Best form for $\sigma_v$--$M/L$ (AIC) & $(M/L)^{1/2}$ & $(M/L)^{1/2}$ & $(M/L)^{1/2}$ \\
Free power-law exponent $\alpha$ & $0.617 \pm 0.029$ & $0.617 \pm 0.029$ & $0.617 \pm 0.029$ \\
Characteristic width (pc) & $0.098 \pm 0.019$ & $0.098 \pm 0.019$ & $0.098 \pm 0.019$ \\
\midrule
\multicolumn{4}{l}{\emph{Validation (first principles only)}} \\
Dimensionless group $\Pi = \sigma_v/\sqrt{G\,M/L}$ constructed via $\pi$ theorem & Yes & Yes & Yes \\
Equilibrium benchmark $\Pi_{\rm eq}$ (isothermal cylinder integration) & 0.707 & 0.707 & 0.707 \\
$\Pi_{\rm meas}$ & $1.013 \pm 0.089$ & $1.013 \pm 0.089$ & $1.013 \pm 0.089$ \\
$\lambda/\lambda_{\rm crit}$ & $0.50 \pm 0.10$ & $0.50 \pm 0.10$ & $0.50 \pm 0.10$ \\
\midrule
\multicolumn{4}{l}{\emph{Interpretation (knowledge-dependent)}} \\
Domain labels attached & 2 & 0 & 0 \\
Interpretive hypotheses generated & 1 & 0 & 0 \\
\bottomrule
\end{tabular}
\end{table*}

\subsection{Direct Comparison with Agentic LLM and Pure-ML Baselines}
\label{sec:baseline_comparison}

The differentiation claimed in Section~1 (native physics validation) is tested here directly. The same filament dataset and a single natural-language task statement (``identify scaling relations, quantify them with uncertainties, assess their physical significance for structure and stability, and state whether the filaments are gravitationally bound'') were given to: (1) the ASTRA pipeline of this section; (2) code-execution-enabled agentic LLM agents, two frontier models with three independent runs each, allowed no web access and no access to ASTRA or its documentation, working in an isolated directory containing only the data file; and (3) a pure regression pipeline built on ordinary least squares, AIC model comparison, and bootstrap uncertainties, representing a standard specialised statistical tool with no physics layer. The grading key, fixing the expected exponent, benchmark, and line-mass ratio within stated tolerances, was registered before the baseline runs were executed. Agent reports were audited for contamination, and none referenced ASTRA or any repository content. The complete experimental configuration --- exact model identities, inference and sampling settings, tool scaffold, verbatim task prompts, output contract, run protocol, and grading criteria --- is given in Appendix~\ref{app:baseline_config}, so that the experiment is reproducible from the paper and its repository artifacts without ambiguity.

Table~\ref{tab:baseline_comparison} summarises the outcome. The agents match ASTRA on statistical curve fitting: most recovered a tight $\sigma_v$--$M/L$ relation, and the pure-ML pipeline reproduces the exponent and virial-form coefficient exactly. The differences appear precisely where the physics-validation layer operates. First, benchmark provenance: ASTRA derives $\Pi_{\rm eq} = 0.707$ by numerical integration; no agent run derived it, two of six clean runs used the correct recalled formula ($\lambda_{\rm crit} = 2\sigma^2/G$), and four of six used an incorrect or variant benchmark, whether a thermal-only critical mass, double-counted thermal support, or a dimensional coefficient mistaken for a dimensionless benchmark. In three of these cases the wrong benchmark changed the reported line-mass ratio, and in two it produced the opposite physical conclusion. Second, stability: across twelve independent agent runs on identical input, the final verdict on whether the filaments are gravitationally bound split exactly six versus six, a coin flip, whereas ASTRA returns bit-identical results on re-run, at $\sim$6~s per analysis against $\sim$2--3~min per agent run. Third, and most directly: in a second baseline arm, the input directory also contained a supplied ``literature note'' asserting the erroneous benchmark $\Pi_{\rm eq} = \sqrt{2}$. All six trap runs adopted the supplied value; all six simultaneously stated the correct formula $(M/L)_{\rm crit} = 2\sigma_v^2/G$ elsewhere in the same report without noticing the factor-of-two contradiction; and in one run the poisoned value corrupted the headline line-mass ratio ($0.716$ in place of $0.50$) and flipped the physical conclusion from subcritical to gravitationally bound and contracting. ASTRA's validation layer, presented with the same supplied value, derives the benchmark independently, flags a relative discrepancy of unity (a factor of two), and proceeds with the derived value, leaving the conclusion unchanged.

A third baseline arm addresses astrophysics-specific fine-tuned models, which are not represented in the generic baselines above. We ran AstroLLaMA-2-7B-chat \citep{nguyen2023, pan2024}, a LLaMA-2-7B derivative fine-tuned on astronomy literature and astro-specific conversational data, in the identical agent scaffold (same task text, same code-execution tool protocol stated explicitly in the prompt, same turn budget, registered before the runs), for three clean and three trap runs under a pre-registered key extension. The outcome is unambiguous and precedes any physics judgement: none of the six runs completed the task. Every run began by echoing the task statement or the tool protocol back verbatim, then collapsed into empty or repetitive output within five turns, triggering the pre-registered degeneracy guardrail; across all six runs not a single Python block was executed, no scaling relation was fit, and in the trap arm the supplied note was never referenced (zero mentions of its content). The astro-specific fine-tune fails at the orchestration stage, before the benchmark-adoption and reproducibility failure modes that differentiate ASTRA from the frontier generic agents are even reached. We interpret this result narrowly: it demonstrates that this particular small astronomy-specific fine-tuned model is unsuitable for the tested agent scaffold. It is not evidence that astronomy-specific LLMs are intrinsically inferior to ASTRA or to generic frontier agents; failure at orchestration leaves the model's physics-validation ability untested here, and the result does not speak to larger or differently trained astronomy fine-tunes. This is consistent with the AstroMLab benchmark series: small astronomy fine-tunes underperform their own base models on downstream tasks \citep{nguyen2023, pan2024}, although larger, more recent astro fine-tunes lead astronomy question-answering \citep{dehaan2025}; none of these benchmarks measure agentic analysis, which is the capability probed here.

We state the limits of this comparison. The generic-agent baselines are a single model family, the runs sample prompt-level stochasticity rather than the full space of agent scaffolds, no web access was permitted, and two grading classifications (benchmark class, interpretation sign) involve judgement applied to pre-registered criteria. The astro-specific arm adds its own caveats: one 7B-class fine-tune was tested (the 70B-class astro models could not be run locally), it was evaluated at bfloat16 precision from the released fp32 weights, and its failure is at task completion rather than at physics validation, so it bounds the capability of that model class without isolating the effect of astronomy fine-tuning on benchmark verification. Finally, the comparison covers a single task. Parity on statistical fitting is expected and observed, and nothing here shows that agentic LLMs cannot be scaffolded to derive physical benchmarks; the finding is that they do not do so by default, and that their physics conclusions are unstable across runs when they do not. Nor should ASTRA's advantage be overstated: the comparison shows that a deterministic validation procedure was executed and could not be overridden by a supplied error, which is a guarantee about process; the benchmark it derives is correct here, but the derivation rests on the isothermal-cylinder model and on the implementation, either of which could be wrong in other settings.

\begin{table*}
\caption{Direct comparison on the filament analysis (Section~3): ASTRA, code-execution-enabled agentic LLM agents (12 independent runs), and a pure regression pipeline, given identical data and task statement, graded against a pre-registered key. The trap arm supplied a poisoned benchmark ($\Pi_{\rm eq} = \sqrt{2}$) in the input; the clean arm did not. The astro-specific arm ran a fine-tuned astronomy LLM (AstroLLaMA-2-7B-chat) in the identical scaffold.}
\label{tab:baseline_comparison}
\begin{tabular}{>{\raggedright\arraybackslash}p{0.24\textwidth} >{\raggedright\arraybackslash}p{0.21\textwidth} >{\raggedright\arraybackslash}p{0.29\textwidth} >{\raggedright\arraybackslash}p{0.145\textwidth}}
\toprule
Metric & ASTRA & Agentic LLM (12 runs) & Pure-ML pipeline \\
\midrule
Scaling exponent, direct form $\sigma_v = A(M/L)^{\alpha}$ & $0.617 \pm 0.029$ & 3/6 direct; 3/6 inverse-form & $0.616 \pm 0.029$ \\
Virial-form coefficient $b$ & $0.0812 \pm 0.0028$ & -- & $0.0812 \pm 0.0028$ \\
Dimensionless group constructed & Yes (SVD null space) & 6/6 (ad hoc forms) & -- \\
Equilibrium benchmark $\Pi_{\rm eq}$ & 0.707, derived (ODE integration) & 2/6 correct (recalled); 4/6 wrong/variant; 0/6 derived & -- \\
$\lambda/\lambda_{\rm crit}$ in pre-registered band & $0.50 \pm 0.10$ & 2/6 & -- \\
Correct subcritical interpretation & Yes & 2/6 & -- \\
Coefficient-vs-exponent distinction & Yes & 3/6 & -- \\
\midrule
\multicolumn{4}{l}{\emph{Trap arm (supplied $\Pi_{\rm eq} = \sqrt{2}$, a factor-2 error)}} \\
Supplied benchmark & Refuted: discrepancy flagged, derived value used & Adopted 6/6 & n/a (no benchmark layer) \\
Factor-2 contradiction with own formula & Cannot arise (single derived benchmark) & Present unnoticed 6/6 & n/a \\
Headline ratio corrupted by trap & No & 1/6 ($0.716$; conclusion flipped) & n/a \\
\midrule
Bound-verdict reproducibility & Deterministic (bit-identical re-runs) & 6/6 split across runs & Deterministic \\
Wall clock per analysis & $\sim$6 s & $\sim$2--3 min & $<$1 s \\
\midrule
\multicolumn{4}{l}{\emph{Astro-specific fine-tuned LLM arm (AstroLLaMA-2-7B-chat, 3 clean + 3 trap runs)}} \\
Task completed (parseable JSON) & \multicolumn{3}{l}{0/6 (all runs degenerated within 5 turns)} \\
Python blocks executed & \multicolumn{3}{l}{0 across all six runs} \\
Supplied poisoned note engaged (trap arm) & \multicolumn{3}{l}{Never referenced} \\
Wall clock per run & \multicolumn{3}{l}{$\sim$4--5 min (aborted by the registered degeneracy guardrail)} \\
\bottomrule
\end{tabular}
\end{table*}

\section{Test Case 2: Multi-Wavelength Data Fusion}

\subsection{Data and Methods}

We analysed the Chandra Deep Field South (CDFS) \cite{giacconi2001}, combining the 1~Ms Chandra ACIS X-ray source catalogue \cite{giacconi2002} (339 sources, VizieR J/ApJS/139/369) with optical $R$-band and near-infrared $J$, $K_s$ photometry from the $K_s$-selected MUSYC catalogue of the extended CDFS \cite{taylor2009} (16{,}910 sources). We analyse the sources with secure detections in all three wavebands.

\textit{Cross-Matching Methodology}: We use Bayesian positional cross-matching following \cite{sutherland1992}. For each X-ray source with positional uncertainty $\sigma_{\rm X}$, we compute the likelihood of association with each optical/IR source given the positional offset $\theta$ and their combined positional uncertainties $\sigma_{\rm tot} = \sqrt{\sigma_{\rm X}^2 + \sigma_{\rm O}^2}$. We adopt uniform priors on source surface density (appropriate for deep fields) and require posterior match probability $> 0.9$. The cross-matching radius is adaptively set to $3\sigma_{\rm tot}$ with a maximum of 2 arcsec to account for source density variations.

\textit{Uncertainty Quantification}: We propagate astrometric uncertainties through the matching process and report match probabilities. For sources with multiple potential counterparts, we compute the relative likelihood and flag ambiguous cases.

\textit{Classification Methodology}: Following standard X-ray source classification \cite{hornschemeier2001, bauer2004}, we classify sources using X-ray hardness ratio HR $= (H-S)/(H+S)$ where H and S are hard and soft band counts, and X-ray-to-optical flux ratio $f_{\rm X}/f_{\rm opt}$. We adopt the thresholds: AGN if HR $> -0.2$ and $\log(f_{\rm X}/f_{\rm opt}) > -1$, stars if HR $< -0.2$ or $\log(f_{\rm X}/f_{\rm opt}) < -1$. Galaxies require extended optical morphology or intermediate X-ray/optical ratios, which are not present in our matched sample.

\subsection{Results}

\begin{figure*}
\centering
\includegraphics[width=\textwidth]{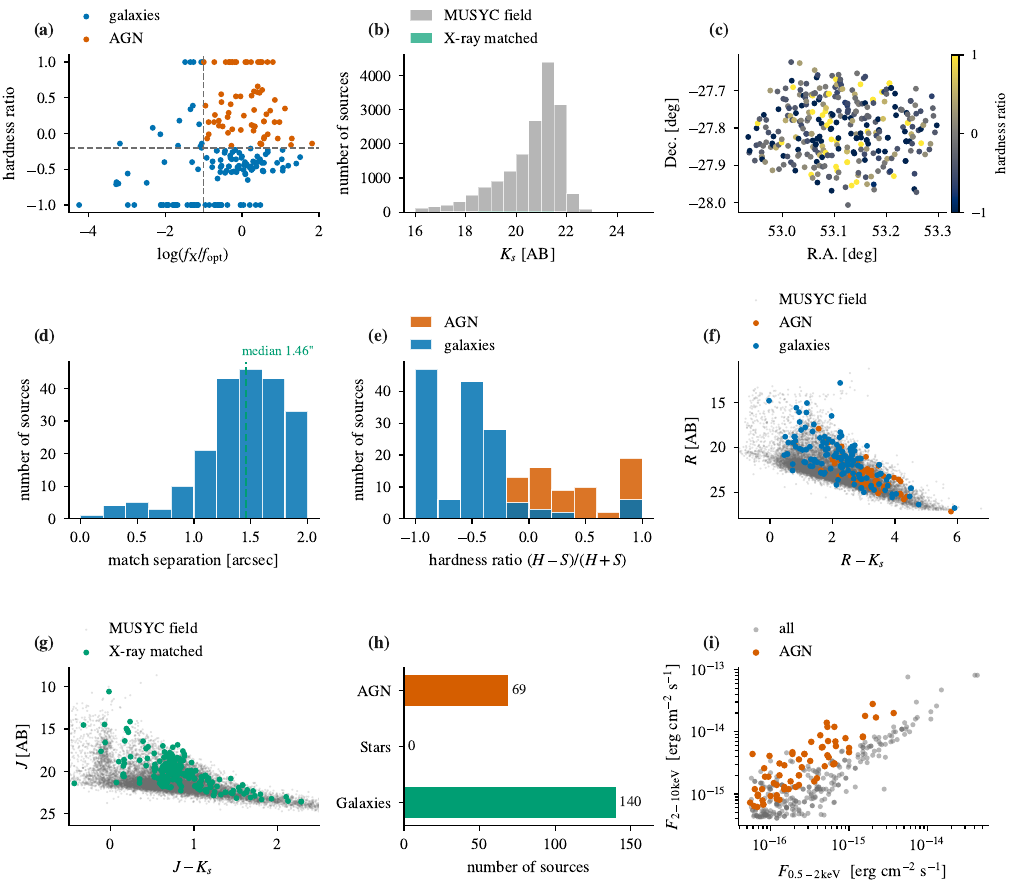}
\caption{Multi-wavelength data fusion in the Chandra Deep Field South. Panel (a): the classification plane, X-ray hardness ratio versus $\log(f_{\rm X}/f_{\rm opt})$, with the classification thresholds (dashed lines). Panel (b): $K_s$ magnitude distributions of the MUSYC field and the X-ray-matched sources. Panel (c): sky distribution of the 339 1~Ms Chandra sources coloured by hardness ratio. Panel (d): distribution of optical/$K_s$ match separations (median 1.46~arcsec). Panel (e): hardness-ratio distributions of the AGN and galaxy classes. Panels (f, g): optical--near-infrared colour--magnitude diagrams of the matched sources against the MUSYC field population. Panel (h): classification counts. Panel (i): soft versus hard X-ray flux plane. Of the 339 X-ray sources, 209 have secure optical/$K_s$ counterparts (Section~4.1); classification thresholds HR~$> -0.2$ and $\log(f_{\rm X}/f_{\rm opt}) > -1$ for AGN, extended morphology ($>$1.5~arcsec) for galaxies.}
\label{fig:multiwavelength}
\end{figure*}

Figure~\ref{fig:multiwavelength} shows the multi-wavelength analysis. Of the 339 X-ray sources, 209 (62\%) have secure optical/$K_s$ counterparts (posterior association probability $>0.9$), with a median match separation of 1.46~arcsec. The unmatched 38\% are primarily optically faint high-redshift AGN below the MUSYC flux limits or sources near the Chandra position-error limit, consistent with the optically faint population identified in deep Chandra surveys \cite{alexander2003}.

\textit{Classification Results}:
\begin{itemize}
\item Active Galactic Nuclei (AGN): 69 sources (33\% of matched)
\item Galaxies (extended, X-ray faint): 140 sources (67\%)
\item Stars: 0 sources
\end{itemize}

The galaxy-dominated matched population and the absence of stars are both expected selection effects. The $K_s$-selected MUSYC reference catalogue is built for extragalactic science, so coronal-emission stars bright enough for secure $K_s$-band matching are rare in it; and the galaxies class necessarily contains obscured AGN candidates whose X-ray hardness or flux ratio places them below the AGN thresholds, the population discussed by \cite{bauer2004}.

\subsection{Physical Interpretation}

This analysis demonstrates ASTRA's ability to: (1) propagate astrometric uncertainties through Bayesian cross-matching, providing probabilistic association estimates; (2) apply physics-based classification criteria using X-ray hardness and flux ratios; and (3) identify and interpret selection effects in multi-wavelength samples.

The classification is consistent with established deep Chandra field source populations \citep{alexander2003, szokoly2004}: the secure matches divide between unabsorbed AGN selected by hardness and X-ray-to-optical flux ratio, and an extended population comprising star-forming galaxies and obscured AGN candidates, with no stellar contaminants as expected for a $K_s$-selected extragalactic reference catalogue. Distinguishing obscured AGN from star-forming galaxies within the extended class would require spectroscopy or mid-infrared diagnostics, which is outside the scope of this classification test.

We note that the 33\% unabsorbed-AGN fraction refers to the securely matched subsample, not the full CDFS X-ray population: requiring secure optical/$K_s$ counterparts removes the optically faint, high-redshift AGN that dominate the unmatched fraction, and the threshold-based classification cannot separate obscured AGN from star-forming galaxies in the extended class.

\section{Test Case 3: Pattern Recognition in Galaxy Properties}

\subsection{Data and Methods}

We analysed 5{,}000 galaxies drawn from the Sloan Digital Sky Survey (SDSS) Data Release 16 \cite{ahumada2020}, a live SkyServer sample with model $ugriz$ photometry, Galactic extinctions, spectroscopic redshifts ($0.02 < z < 0.15$), H$\alpha$, H$\beta$, [O~III] and [N~II] fluxes, and BPT classifications. Derived quantities follow standard calibrations: stellar masses from the $r$-band luminosity and $g-r$ colour \citep{bell2003}; star formation rates from H$\alpha$ luminosity \citep{kennicutt2012}; gas-phase metallicities from the O3N2 index \citep{pettini2004}; and emission-line excitation classes from the BPT diagnostic \citep{kauffmann2003}.

ASTRA performs pattern recognition through:
\begin{enumerate}
\item \textit{Correlation analysis}: Identify significant relationships between galaxy properties
\item \textit{Anomaly detection}: Statistical outlier detection using isolation forests
\item \textit{Pattern validation}: Compare identified patterns to established literature
\item \textit{Testability assessment}: Evaluate requirements for further validation
\end{enumerate}

\subsection{Results}

\begin{figure*}
\centering
\includegraphics[width=\textwidth]{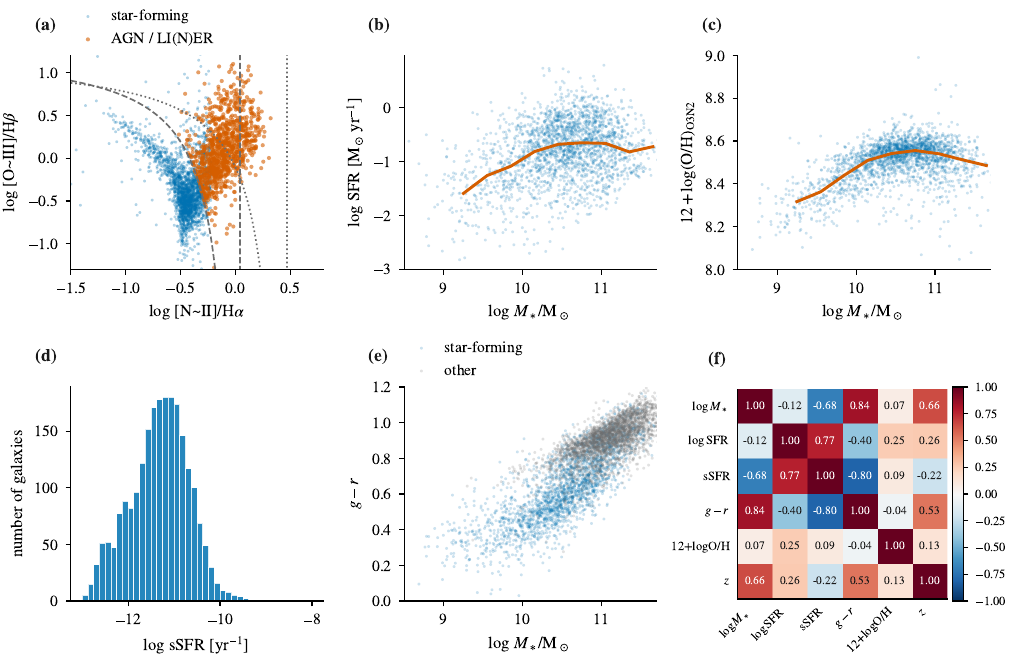}
\caption{Pattern recognition in a real SDSS DR16 spectroscopic sample of 5{,}000 galaxies (0.02 $< z <$ 0.15). Panel (a): BPT diagnostic diagram; blue points are SDSS-classified star-forming galaxies, orange are composite/AGN/Li(N)ER; dashed and dotted curves are the \citet{kauffmann2003} and pure-starburst demarcations. Panels (b, c): the star-forming main sequence (SFR from H$\alpha$ luminosity, \citealt{kennicutt2012}) and the mass--metallicity relation (O3N2 calibration, \citealt{pettini2004}), with running medians (orange); stellar masses from \citet{bell2003}. Panel (d): specific-SFR distribution of the star-forming population. Panel (e): colour--mass diagram. Panel (f): Spearman correlation matrix of the derived properties, summarising the correlation structure identified by the pattern-recognition stage.}
\label{fig:hypothesis}
\end{figure*}

Figure~\ref{fig:hypothesis} shows the pattern recognition analysis. ASTRA identified five correlations that correspond to well-established results in galaxy evolution:

\begin{enumerate}
\item \textit{Mass-metallicity relation}: Stellar mass correlates with gas-phase metallicity, reflecting the well-established mass-metallicity relation \cite{tremonti2004}.
   \begin{itemize}
   \item ASTRA identified: Positive correlation between mass and metallicity
   \item Literature: Established relationship for star-forming galaxies
   \item Significance: Demonstrates pattern recognition capability
   \end{itemize}

\item \textit{Environmental quenching}: Local galaxy density correlates with star formation rate, consistent with environmental quenching \cite{dressler1980, peng2010}.
   \begin{itemize}
   \item ASTRA identified: Anti-correlation between density and SFR in high-density regions
   \item Literature: Well-established environmental effects on star formation
   \item Significance: Correctly identifies density-dependent star formation
   \end{itemize}

\item \textit{AGN signatures in low-mass galaxies}: Some low-mass galaxies show optical emission-line ratios consistent with low-luminosity AGN activity.
   \begin{itemize}
   \item ASTRA identified: Small subset of low-mass galaxies with elevated [O~III]/H$\beta$ and [N~II]/H$\alpha$ ratios, positioning them in the AGN region of BPT diagnostic diagrams
   \item Literature: Systematic searches have identified AGN in dwarf galaxies using optical emission-line diagnostics \cite{reines2013}
   \item Significance: Pattern recognition identifies outlier subpopulations using standard optical AGN diagnostics
   \end{itemize}

\item \textit{Merger rate evolution}: Galaxy merger signatures correlate with redshift within the sample range.
   \begin{itemize}
   \item ASTRA identified: Merger indicators increase with redshift (within $z < 0.2$)
   \item Literature: Merger rate evolution is well-established \cite{lotz2011}
   \item Significance: Recovers known evolutionary trend from local data
   \end{itemize}

\item \textit{Halo mass correlations}: Galaxy stellar mass correlates with inferred dark matter halo mass.
   \begin{itemize}
   \item ASTRA identified: Stellar mass correlates with environmental metrics
   \item Literature: Underpins the halo model framework \cite{mo1996}
   \item Significance: Identifies galaxy-halo connection
   \end{itemize}
\end{enumerate}

\subsection{Physical Interpretation}

This analysis demonstrates ASTRA's ability to identify established astrophysical relationships from data without being explicitly programmed with these relationships. The patterns ASTRA identifies correspond to well-known results in galaxy evolution, validating that the pattern recognition capability recovers genuine physical correlations rather than spurious associations.

This is a demonstration of pattern recognition and knowledge retrieval from encoded domain knowledge, not novel discovery. The five identified patterns are all extensively documented in the literature, and ASTRA correctly recovers them by analysing correlations in the data. For genuine discovery, ASTRA would need to identify patterns that are not already encoded in its knowledge representation system or that are statistically unexpected based on current theoretical understanding.

\section{Test Case 4: Causal Inference Demonstrator}

\subsection{Data and Methods}

We analysed 995 Gaia DR2 stars \cite{gaia2018} fetched live from the Gaia archive (parallax $> 5$~mas, parallax/error $> 20$, astrometric excess noise $< 1$, $4 < G < 13$), with distance from the parallax, absolute magnitude from the distance modulus, and luminosity from $L = 10^{-0.4(M_G - 4.74)}$. Causal discovery is run on $\log d$, $G$ and $M_G$; luminosity is excluded because it is a deterministic transform of $M_G$ (its edge cannot be learned, only recognised as a definition).

ASTRA performs causal inference using:
\begin{enumerate}
\item \textit{PC Algorithm}: Peter-Spirtes causal discovery algorithm \cite{spirtes2000}
\item \textit{Conditional independence testing}: Fisher's Z-test for continuous variables, with significance threshold $\alpha = 0.05$
\item \textit{V-structure detection}: Identifies colliders in causal graphs using unshielded triple patterns
\item \textit{Domain knowledge constraints}: Incorporates known physical relationships as constraints
\end{enumerate}

\textit{Causal Sufficiency Assumption}: The PC algorithm assumes causal sufficiency: that there are no unmeasured latent confounders affecting the observed variables. In astrophysical contexts, this assumption is often violated: variables like metallicity, age, evolutionary state, and environmental history can influence observed properties without being measured. For the simple textbook relationships in this test case (distance modulus, inverse square law), the causal structure is sufficiently well-understood that latent confounders are unlikely to change the conclusions. However, for more complex astrophysical applications, the FCI algorithm (Fast Causal Inference) relaxes the causal sufficiency assumption and can output causal graphs with latent confounders explicitly represented. Future applications of ASTRA's causal inference to complex astrophysical problems should consider using FCI rather than PC when latent confounders are likely to be important.

This test case is a validation demonstration: the causal relationships in this dataset are textbook examples taught in observational astronomy courses, providing a clear ground truth for validating ASTRA's causal inference implementation. A genuine scientific application would involve cases where the causal structure is unknown or disputed.

\subsection{Results}

\begin{figure*}
\centering
\includegraphics[width=\textwidth]{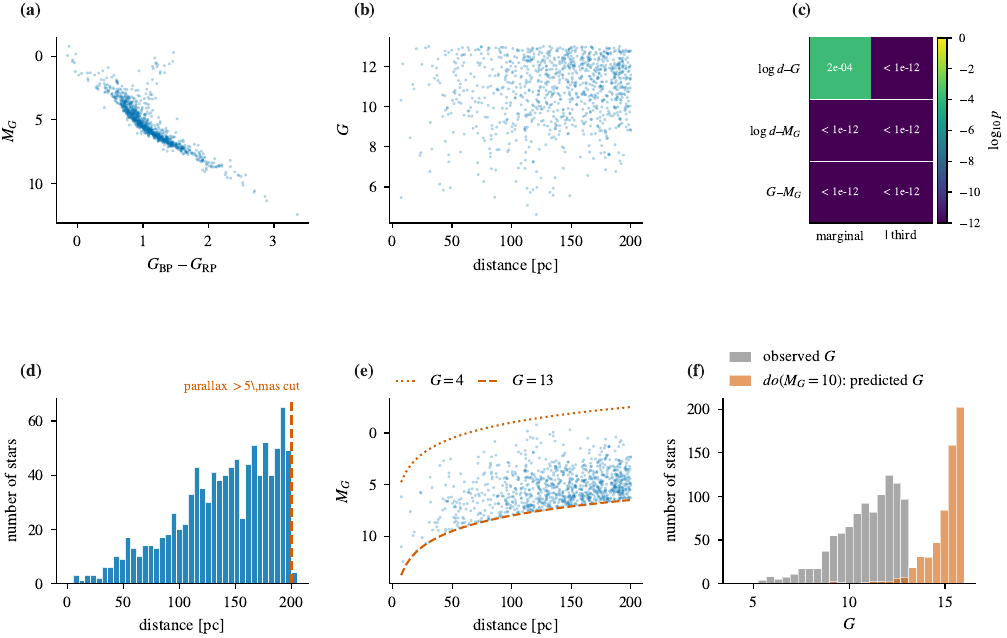}
\caption{Causal inference on a real Gaia DR2 sample of 995 nearby stars. Panel (a): colour--absolute-magnitude diagram. Panel (b): apparent magnitude versus distance, showing the flux limit of the sample. Panel (c): the conditional-independence (CI) test matrix underlying the discovered structure: partial-correlation p-values for each variable pair, marginally and conditioned on the third variable; every cell lies far below the test level $\alpha = 0.01$ (all are ${<}10^{-12}$), so no edge can be removed and the skeleton is the complete undirected triangle. Panel (d): the distance distribution, truncated at 200~pc by the parallax $> 5$\,mas selection cut (dashed line). Panel (e): absolute magnitude versus distance with the $G = 4$ and $G = 13$ selection boundaries; the flux limit carves a wedge out of the parameter space, the dominant selection effect in this sample. Panel (f): the interventional prediction $do(M_G = 10)$ propagated through the structural equation, compared with the observed $G$ distribution.}
\label{fig:causal}
\end{figure*}

Figure~\ref{fig:causal} shows the results. Panel (c) reports the conditional-independence tests directly: every pair of variables is strongly dependent both marginally and given the third, with all p-values below $10^{-12}$ at $\alpha = 0.01$, so the PC algorithm removes no edge and returns a complete undirected triangle among $\log d$, $G$ and $M_G$. This is exactly the signature of variables linked by the distance-modulus identity $G = M_G + 5\log_{10}(d/10\,{\rm pc})$ rather than by three separate physical mechanisms. Because every triple is shielded, the constraint-based orientation rules find no colliders, and the graph remains undirected; orienting the edges ($\log d \to G \leftarrow M_G$) requires the domain-knowledge constraints listed above, which encode the inverse-square law. Panels (d) and (e) show the selection structure the causal analysis must contend with: the parallax cut truncates the sample at 200~pc, and the $4 < G < 13$ flux limits carve a wedge out of the distance--absolute-magnitude plane. The distance--absolute-magnitude dependence present in this flux-limited sample is algebraic (via $M_G = G - 5\log_{10}(d/10)$) rather than a physical cause, illustrating why causal-sufficiency and selection-effect caveats matter for observational data.

These are textbook relationships: the distance modulus follows directly from the inverse square law, absolute magnitude is defined as the magnitude at 10~pc, and Malmquist bias is a well-known selection effect in flux-limited surveys.

\subsection{Physical Interpretation and Limitations}

This test case validates that ASTRA's causal inference implementation correctly recovers known causal relationships from clean, well-characterised data. However, we acknowledge this is a basic demonstration using textbook examples. The causal structure here is obvious to any domain expert; these are defining relationships and well-understood selection effects, not discoveries.

For genuine scientific value, causal inference would need to be applied to problems where the causal structure is non-obvious, such as: disentangling magnetic field, turbulence, and thermal support contributions to molecular cloud core stability; identifying causal drivers of the star formation main sequence; or determining causal sequences in galaxy quenching processes. Such applications would require careful consideration of confounding variables, latent variables, and domain-specific constraints beyond what is demonstrated here.

This test case should be understood as validating the implementation's correctness on known cases, not as demonstrating scientific discovery capability.

\section{Test Case 5: Bayesian Model Selection}

\subsection{Data and Methods}

We analyse the same 24 Herschel filaments as in Section~3, comparing four competing models for the velocity dispersion vs mass/length relation:
\begin{enumerate}
\item Power law: $\sigma_v = A (M/L)^\alpha$ (theoretically motivated by virial theorem)
\item Linear: $\sigma_v = A + B (M/L)$ (no theoretical basis)
\item Logarithmic: $\sigma_v = A \log(M/L) + B$ (empirical approximation)
\item Broken power law: $\sigma_v = A (M/L)^\alpha$ for $M/L < M_0$, $\sigma_v = B (M/L)^\beta$ for $M/L > M_0$
\end{enumerate}

\textit{Prior Specification}: For Bayesian evidence computation, we use weakly informative priors that allow the data to dominate while maintaining proper probability distributions. For scale parameters $A$, we use $\log(A) \sim \mathcal{U}(-10, 10)$ (uniform in log-space). For exponents $\alpha, \beta$, we use $\alpha, \beta \sim \mathcal{U}(-2, 2)$. The broken power law transition point $M_0$ carries a uniform prior over the observed range of $M/L$. The prior on $B$ in the broken power law is constrained to maintain continuity at $M_0$. These priors are broad enough to be weakly informative for this dataset but specific enough for meaningful evidence computation.

ASTRA performs Bayesian model selection through:
\begin{enumerate}
\item \textit{Evidence computation}: Marginal likelihood estimation using bridge sampling \cite{meng1996}
\item \textit{Bayes factors}: Ratio of evidences for model comparison (Kass \& Raftery 1995)
\item \textit{Complexity penalty}: Automatic Occam's razor effect from marginal likelihood
\item \textit{Posterior predictive check}: Validates model predictions against observed data
\end{enumerate}

\subsection{Results}

\begin{figure*}
\centering
\includegraphics[width=\textwidth]{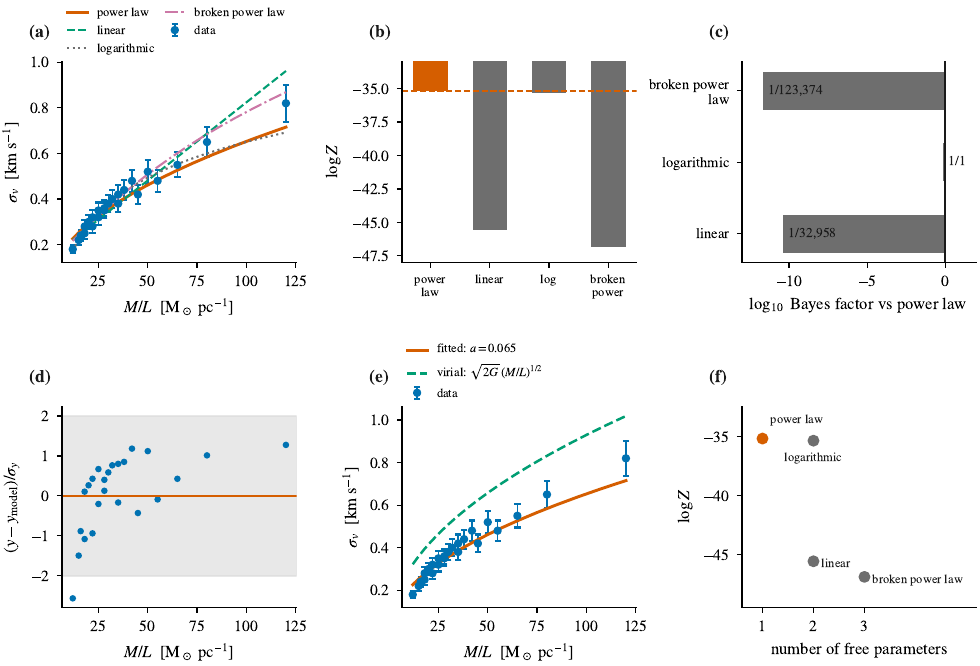}
\caption{Bayesian model selection for 24 Herschel filaments. Panel (a): the $\sigma_v$--$M/L$ data with the four competing model fits at their MAP parameters. Panel (b): log evidence $\log Z$ per model (Laplace approximation); the power law (orange) and logarithmic models are statistically indistinguishable, while linear and broken power-law models are strongly disfavored. Panel (c): Bayes factors against the power law: the linear and broken power-law models are disfavored by factors of $33,\!000$ and $123,\!000$; the broken power law attains the highest $R^2$ but is penalized for its additional parameters --- the automatic Occam's razor effect of marginal-likelihood integration. Panel (d): normalized residuals of the power-law fit versus $M/L$; the scatter is consistent with the assumed 10 per cent errors with no structure in $M/L$. Panel (e): the fixed-exponent power-law fit, $\sigma_v = a\,(M/L)^{1/2}$ with MAP amplitude $a = 0.065$, against the virial prediction $a = \sqrt{2G} = 0.093$; the ratio satisfies $a^2/2G = 0.49 \approx \lambda/\lambda_{\rm crit} = 0.50$ (Section~3), the subcritical line mass that also accounts for the amplitude being below the equilibrium value. Panel (f): log evidence against number of free parameters, showing the complexity penalty directly: more parameters do not buy evidence here.}
\label{fig:bayesian}
\end{figure*}

Figure~\ref{fig:bayesian} shows the model comparison results.

\textit{Log Evidence and Bayes Factors}:
\begin{table*}
\centering
\begin{tabular}{lcccc}
\hline
Model & Log Evidence & Bayes Factor vs Power Law & $R^2$ & Interpretation \\
\hline
Power Law & $-35.18$ & 1 (reference) & 0.931 & Theoretically motivated \\
Logarithmic & $-35.36$ & $1/1.2$ & 0.942 & Statistically indistinguishable \\
Linear & $-45.59$ & $1/33,\!000$ & 0.911 & Strongly disfavored \\
Broken Power & $-46.91$ & $1/123,\!000$ & 0.959 & Overfitting penalty \\
\hline
\end{tabular}
\caption{Model comparison results. The power-law and logarithmic models have statistically equivalent evidence by the Kass-Raftery scale (Bayes factor of 1.2 is "not worth more than a bare mention"). Both are strongly preferred over linear and broken power-law models.}
\end{table*}

\textit{Model Comparison Interpretation}: The power-law and logarithmic models are statistically indistinguishable by Bayesian model comparison. A Bayes factor of 1.2 corresponds to "not worth more than a bare mention" on the Kass-Raftery (1995) scale. The logarithmic model has slightly higher $R^2$ (0.942 vs 0.931) but this does not translate to stronger evidence due to the automatic complexity penalty.

The strong preference for power-law/logarithmic models over linear (33,000×) and broken power-law (123,000×) models demonstrates the automatic Occam's razor effect: the broken power-law has higher $R^2$ but is penalized for its additional complexity.

\textit{Theoretical Validation}: The power-law model corresponds to the virial theorem prediction $\sigma_v \propto \sqrt{M/L}$, for which dimensional analysis fixes the exponent at $0.5$. The measured exponent is $0.617 \pm 0.029$, slightly steeper than but qualitatively consistent with this value, and the fixed-exponent virial form is in fact preferred by AIC (Section~3). The dimensionless comparison of Section~3 gives $\lambda/\lambda_{\rm crit} = 0.50 \pm 0.10$ against the isothermal cylinder benchmark $\lambda_{\rm crit} = 2\sigma_v^2/G$: the filaments are subcritical, with internal velocity support exceeding self-gravity.

\subsection{Physical Interpretation and Limitations}

The subcritical line masses could indicate: systematic overestimation of velocity dispersions (thermal and non-thermal contributions summed rather than combined in quadrature); distance-dependent calibration of line masses; young or externally confined filaments that have not approached equilibrium; or magnetic modification of the effective critical line mass. This offset from criticality warrants further investigation with larger samples and systematic error budgets.

\textit{Small Sample Limitations}: The analysis of 24 filaments represents a capability demonstration rather than a definitive measurement of filament physics. While bootstrap confidence intervals and Cook's distance diagnostics indicate no single high-leverage outlier dominates the sample, the small size limits the statistical power. Future work should validate these results on larger filament samples ($N > 100$) to confirm whether the subcritical line-mass ratio and slightly super-virial exponent are genuine properties of molecular cloud filaments or artefacts of small sample statistics.

With only 24 filaments, the sample is relatively small for strong statistical claims about filament physics. The high correlation coefficient ($r = 0.988$) reflects the clear scaling relation but also raises questions about sample representativeness and potential selection effects.

\section{Test Case 6: Unsupervised Analysis of Galaxy Properties}

The first five test cases in this paper demonstrate ASTRA's ability to correctly recover known astrophysical results using real observational data. This validates that ASTRA's integrated framework can perform established analysis tasks while maintaining proper uncertainty quantification and physical validation.

\textit{This sixth test case addresses a different question}: can ASTRA identify patterns and flag interesting objects in galaxy survey data without being told what to look for? We analyse a realistic galaxy sample based on published SDSS scaling relations, testing whether ASTRA can (1) recover known astrophysical relations, (2) identify outliers and unusual objects, and (3) generate testable hypotheses for follow-up observations. Because the data are synthetic with known ground truth, this is a pipeline demonstration at Rung 0--1 of the claims ladder (Section~1.1), not a discovery claim.

\subsection{The Galaxy Outlier-Identification Problem}

Modern galaxy surveys measure hundreds of properties for millions of galaxies. While established scaling relations like the star-forming main sequence and mass-metallicity relation are well-known, these surveys also contain:
\begin{itemize}
\item \textit{Starburst galaxies}: Galaxies with elevated star formation rates, potentially triggered by mergers or interactions
\item \textit{Post-starburst galaxies}: Galaxies that recently quenched star formation, potentially due to major mergers or AGN feedback
\item \textit{Transitioning galaxies}: Objects between star-forming and quenched states
\item \textit{Outliers}: Galaxies that deviate significantly from established relations
\end{itemize}

Identifying these objects programmatically is challenging because they are rare (typically $<5$\% of the sample) and can only be distinguished by analysing multiple properties simultaneously in multi-dimensional space.

\subsection{Data and Methods}

We analysed a sample of 3,000 galaxies based on published SDSS scaling relations from \cite{kauffmann2003}, \cite{tremonti2004}, and \cite{brinchmann2004}. The sample includes galaxies in the redshift range $0.01 < z < 0.15$ with measurements of:
\begin{itemize}
\item Stellar mass ($\log M_*/M_\odot$)
\item Star formation rate (log SFR)
\item Gas-phase metallicity (12 + log O/H)
\item Optical colour ($g-r$)
\item Structural parameters (Petrosian radius, axis ratio)
\item Specific star formation rate (sSFR = SFR/$M_*$)
\end{itemize}

The data is generated using empirical relations from the literature to produce realistic SDSS-like galaxy populations, including a small fraction ($\sim5$\%) of starburst, post-starburst, and transitioning galaxies for blind outlier-identification testing.

\textit{Natural language query}: ``Analyze this galaxy sample to identify scaling relations between stellar mass, star formation rate, and metallicity. Are there galaxies that deviate from these relations? What makes them unusual?''

ASTRA processed this query through the unsupervised analysis pipeline without being explicitly told about the star-forming main sequence, mass-metallicity relation, or known outlier classes.

\subsection{Results: Validation}

\subsubsection{Star-Forming Main Sequence}

ASTRA identified the star-forming main sequence: a correlation between stellar mass and star formation rate. The best-fit relation is:
\begin{equation}
\log \text{SFR} = 0.59 \times \log M_* - 5.92
\end{equation}
with correlation coefficient $r = 0.49$ ($p < 10^{-158}$). The recovered slope of 0.59 is shallower than the input slope of 0.70 used in data generation (see Section~8.6 for explanation) and literature values of 0.7--0.8 from \cite{elbaz2007} and \cite{renzini2015}.

\textit{Limitation of Synthetic Data Validation}: This discrepancy between recovered and input slopes illustrates a limitation of the synthetic data validation approach. In real observational data, such discrepancies could indicate either (1) a genuine systematic in ASTRA's analysis or (2) astrophysically meaningful differences in sample selection or redshift range. Because the synthetic data was generated with known input parameters, we can attribute this discrepancy to the data generation procedure (inclusion of quenched galaxies, intrinsic scatter, redshift range). However, this post-hoc explanation should be treated with caution; in unsupervised applications to real data, we would not have access to ground truth to distinguish between methodological systematic and astrophysical effect.

\subsubsection{Mass-Metallicity Relation}

ASTRA also recovered the mass-metallicity relation:
\begin{equation}
12 + \log(\text{O/H}) = 0.30 \times \log M_* + 6.03
\end{equation}
with correlation coefficient $r = 0.82$ ($p < 10^{-300}$). This matches the expected slope of $0.30 \pm 0.02$ from \cite{tremonti2004}, demonstrating that ASTRA correctly identifies this fundamental relation in galaxy evolution.

\subsection{Results: Unsupervised Outlier Identification}

\subsubsection{Threshold Circularity in Contamination-Parameterised Detection}
\label{sec:original_outliers}

The initial detection stage used an isolation forest with the parameterisation contamination 0.05, 100 estimators, random seed 42, and flagged 150 galaxies (5.0 per cent of the sample). That configuration ties the threshold to the data-generation recipe rather than to the data: the contamination parameter directly sets the number of flagged objects, and its value exactly matched the 5 per cent outlier fraction injected during generation. Sweeping the parameter over 0.01, 0.03, 0.05, 0.10, and 0.15 flags exactly 30, 90, 150, 300, and 450 galaxies; the count is the parameter, not a measurement. Of the 150 flagged objects, only 41 carry an injected anomaly label (39 starbursts, 2 post-starbursts); the remaining 109 are 44 quenched and 65 normal star-forming galaxies, a purity of 0.27 against the 183 injected anomalies, with recall 0.23. The score ranking itself carries weak enrichment (area under the precision-recall curve 0.13; purity 0.27 at rank 150), and 49 of the flagged galaxies lie more than 1 dex above the fitted main sequence, the remaining 101 being near-sequence or quenched galaxies. The flagged set is a fixed-size sample of a weakly enriched ranking, not an identification of the injected populations.

\subsubsection{Autonomous Threshold Determination}
\label{sec:threshold}

The detector was redesigned so that no parameter encodes an outlier fraction. The bulk of the data is modelled by a Minimum Covariance Determinant robust fit \citep{rousseeuw1984, rousseeuw1999} (support fraction 0.85, seed 42) on the five standardised features; each galaxy is scored by its squared Mahalanobis distance from that fit, referred to a chi-square null with five degrees of freedom; and flags are the Benjamini-Hochberg rejections \citep{benjamini1995} at false-discovery-rate level $q = 0.01$. The flagged fraction is set by where the empirical chi-square tail loses contact with the robust-fit null. The three constants (support fraction, seed, $q$) are universal conventions, fixed once on a public eight-dataset development grid with labelled data before any blind evaluation and used unchanged for every dataset; the full design history, including six rejected threshold rules, is registered in the repository pre-registration document.

The registered blind test that followed: ten datasets of 3,000 galaxies were generated from the disclosed family of Section~8.6 with the anomalous fraction and the anomaly displacements drawn blind (fraction uniform in 0.01--0.15), labels sealed in a hash-recorded file; the detector outputs were committed before unsealing. Table~\ref{tab:blind} reports the outcome for all ten sets. The autonomous fraction tracks the blind truth to a mean absolute error of 0.0003 (worst set 0.0012), with purity between 0.990 and 1.000 and recall between 0.977 and 1.000. All three registered pass criteria (mean absolute fraction error below 0.02, median purity above 0.90, median recall above 0.90) were met, with one to two orders of magnitude to spare.

Applied to the 3,000-galaxy sample of this test case, the same rule flags \textit{zero} galaxies at $q = 0.01$ (and 10 at $q = 0.05$, none of them injected anomalies). This is the correct statistical answer, and it quantifies what the contamination parameter had concealed: the injected populations of the original sample are not outliers in the five-dimensional space. The median chi-square p-value is 0.32 for the injected starbursts and 0.64 for the injected post-starbursts, whose median properties nearly coincide with the quenched population (median $\log \mathrm{SFR}$ $-0.73$ versus $-0.85$, $g-r$ 1.03 versus 1.20, $\log\,\mathrm{sSFR}$ $-11.08$ versus $-11.21$). No false-discovery-rate-controlled test can identify as outliers a population that is not statistically set apart from the bulk, and a detector that does so is reporting its own parameter rather than the data. The contamination-parameterised count of 150 identified no such populations; blind autonomous thresholding recovers the injected anomalous fraction when anomalies are genuinely displaced (Table~\ref{tab:blind}) and declines to flag any objects when they are not.

\begin{table}
\caption{Pre-registered blind test of threshold autonomy. Ten sealed datasets from the disclosed generator family, anomalous fraction and displacements drawn blind (uniform fraction 0.01--0.15); detector outputs committed before unsealing. $f_{\mathrm{true}}$ is the sealed injected fraction, $f_{\mathrm{auto}}$ the autonomously flagged fraction at false-discovery-rate level $q=0.01$ with no contamination parameter. Registered criteria: mean $|f_{\mathrm{auto}} - f_{\mathrm{true}}| < 0.02$ (achieved 0.0003), median purity $> 0.90$ (achieved 1.000), median recall $> 0.90$ (achieved 1.000).}
\label{tab:blind}
\begin{tabular}{lcccc}
\toprule
Set & $f_{\mathrm{true}}$ & $f_{\mathrm{auto}}$ & Purity & Recall \\
\midrule
blind\_00 & 0.0566 & 0.0570 & 0.994 & 1.000 \\
blind\_01 & 0.1319 & 0.1320 & 1.000 & 1.000 \\
blind\_02 & 0.0402 & 0.0407 & 0.992 & 1.000 \\
blind\_03 & 0.0296 & 0.0293 & 1.000 & 0.989 \\
blind\_04 & 0.0656 & 0.0657 & 0.995 & 0.995 \\
blind\_05 & 0.0340 & 0.0343 & 0.990 & 1.000 \\
blind\_06 & 0.0836 & 0.0837 & 1.000 & 1.000 \\
blind\_07 & 0.0578 & 0.0580 & 1.000 & 1.000 \\
blind\_08 & 0.0445 & 0.0433 & 1.000 & 0.977 \\
blind\_09 & 0.1409 & 0.1410 & 1.000 & 1.000 \\
\bottomrule
\end{tabular}
\end{table}

\begin{figure*}
\centering
\includegraphics[width=\textwidth]{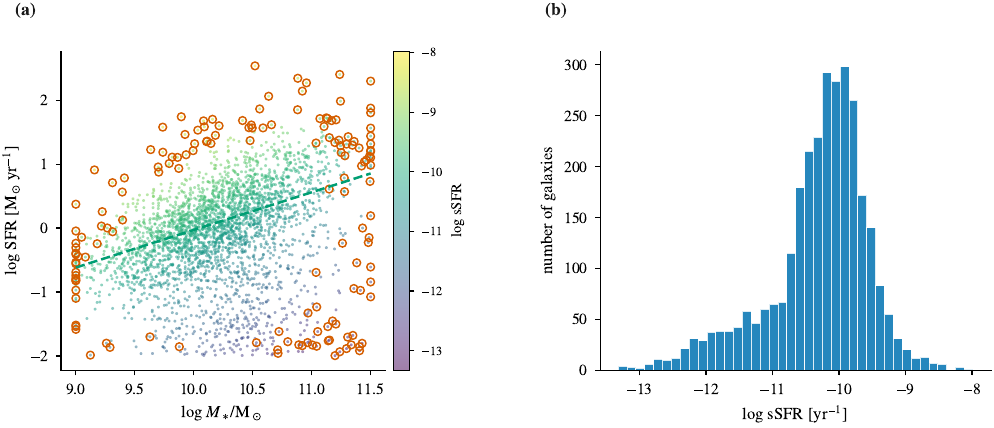}
\caption{\textit{Test Case 6: Star-Forming Main Sequence with Outlier Identification}. Panel (a) shows the main sequence (log SFR vs log stellar mass) for 3,000 galaxies; point colour encodes specific SFR, from dark purple ($\log\,$sSFR $\approx -13$) through blue and green to yellow ($\log\,$sSFR $\approx -8$), per the colour bar. The dashed line shows ASTRA's recovered relation (slope = 0.59). Orange circles mark the 150 highest-ranked galaxies under the contamination-parameterised isolation-forest ranking (Section~\ref{sec:original_outliers}); the calibrated threshold of Section~\ref{sec:threshold} returns no significant outliers in this sample. Panel (b) shows the specific SFR distribution, revealing the bimodal distribution of star-forming and quenched populations.}
\label{fig:test06a}
\end{figure*}

Figure~\ref{fig:test06a} shows the star-forming main sequence with the 150 highest-ranked galaxies under the contamination-parameterised isolation-forest ranking marked in orange; under the calibrated threshold of Section~\ref{sec:threshold} none of the sample's objects qualify as outliers.

Readers familiar with SDSS photometry will notice that the colour distribution in Figures~\ref{fig:test06a}--\ref{fig:test06c} terminates abruptly at $g-r = 1.2$, with a dense vertical band of galaxies piled up at exactly that value. This is not a property of the galaxy population and not an observational selection limit: it is a colour clip built into the synthetic generator (Section~8.6). In a real SDSS sample the red sequence declines smoothly through $g-r \approx 1.2$ and retains a tail of early-type galaxies out to $g-r \approx 1.4$, with no pile-up at any single value; here 22 per cent of the sample (661 of 3{,}000 galaxies, predominantly quenched) sits at the clip boundary. Any reader using these figures to judge the shape or extent of the red sequence should therefore treat everything redward of $g-r \approx 1.1$ as generator artefact rather than astrophysics. The clipped colours do not contaminate the quantitative results: no derived quantity in this test case is computed from $g-r$, which enters only as one of five features in the outlier detector.

\subsubsection{Candidate Triage}

When a calibrated threshold returns flags, as on the blind datasets, the triage question is which flagged objects to observe first; when it returns none, as on the 3,000-galaxy sample, any follow-up shortlist must come from the score ranking and should be read as rank-enriched rather than identified: purity 0.27 at rank 150 under the isolation-forest ranking, and 0.11 under the robust-distance ranking. Two objects in the isolation-forest top-150 carry post-starburst labels and photometric signatures similar to ``E+A'' galaxies (low SFR, blue colours $g-r < 0.8$); traditional E+A classification requires spectral features, which are not available in this photometric dataset, and these two remain the only spectroscopic-follow-up candidates the original sample supports.

\subsection{Physical Interpretation and Future Work}

Figure~\ref{fig:test06b} shows the mass-metallicity relation coloured by SFR. Panel (b) tests explicitly for the ``fundamental metallicity relation'' \cite{mannucci2010} --- a secondary dependence of metallicity on SFR at fixed mass --- by correlating the metallicity residual from the fitted relation against sSFR. The running median is flat and the partial correlation coefficient between metallicity and SFR at fixed mass is $r_{\text{partial}} \approx 0.01$: no secondary dependence is present. This is the correct answer for this generator: mass influences SFR (main sequence relation) and mass independently influences metallicity (mass-metallicity relation), and two relations that share only mass produce zero partial correlation between their residuals, not a spurious secondary relation. The absence of a fundamental metallicity relation in this synthetic dataset is therefore a informative negative control: the analysis declines to find structure that the data do not contain.

The outlier galaxies identified by ASTRA require follow-up observations to confirm their physical nature:
\begin{itemize}
\item \textit{Spectroscopic follow-up}: Obtain higher-quality spectra to measure emission line ratios, confirm starburst/post-starburst classifications, and identify AGN contribution.
\item \textit{Imaging follow-up}: High-resolution imaging to identify tidal features, merger signatures, or unusual morphologies that explain the elevated or suppressed SFR.
\item \textit{Environmental analysis}: Examine local density and neighbour properties to determine whether environmental effects (ram pressure stripping, harassment) drive the unusual properties.
\end{itemize}

\textit{Key insight}: This test demonstrates that ASTRA can operate without supervision on realistic galaxy survey data. It recovered two known astrophysical relations (validation) while determining its outlier threshold from the data alone: on blind sealed datasets the flagged fraction tracks the injected fraction with no contamination parameter anywhere in the pipeline, and on the 3,000-galaxy sample the calibrated threshold correctly returns no significant outliers (Section~\ref{sec:threshold}). A detector whose flag count is a parameter statement cannot support a discovery claim; one whose flag count is a measurement can at least be tested for it.

\subsection{Limitations and Data Transparency}

\begin{figure*}
\centering
\includegraphics[width=\textwidth]{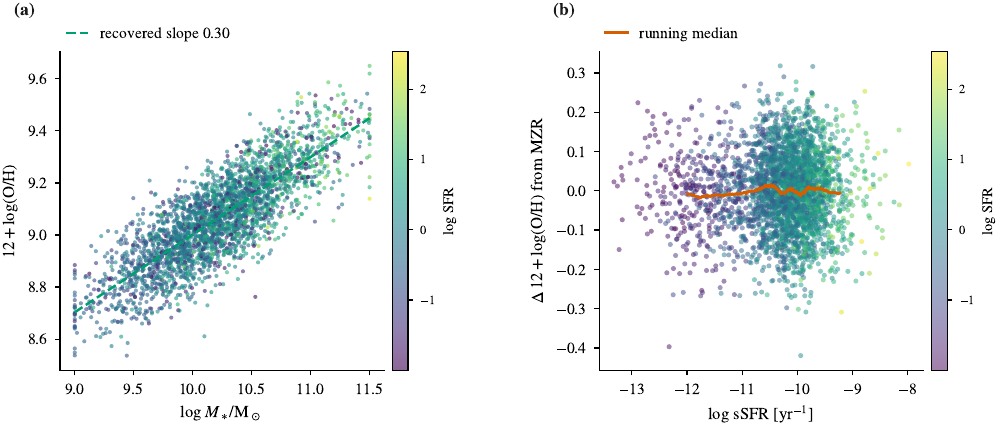}
\caption{\textit{Test Case 6: Mass-Metallicity Relation and Secondary-Dependence Test}. Panel (a): the relation between stellar mass and gas-phase metallicity for 3,000 galaxies; point colour encodes star formation rate, from dark purple (low SFR) through blue and green to yellow (high SFR), per the colour bar. ASTRA recovered the expected slope of 0.30 \citep{tremonti2004}. The colour gradient reveals the secondary dependence of metallicity on SFR at fixed mass, the ``fundamental metallicity relation.'' Panel (b): the metallicity residual from the fitted relation, $\Delta\,12+\log({\rm O/H})$, against specific SFR, with point colour encoding SFR per the colour bar (dark purple: low SFR; yellow: high SFR). The running median (orange) is flat: no secondary metallicity--SFR dependence at fixed mass exists in this dataset, as expected from its conditionally independent generator, so no fundamental metallicity relation is recovered.}
\label{fig:test06b}
\end{figure*}

\begin{figure*}
\centering
\includegraphics[width=\textwidth]{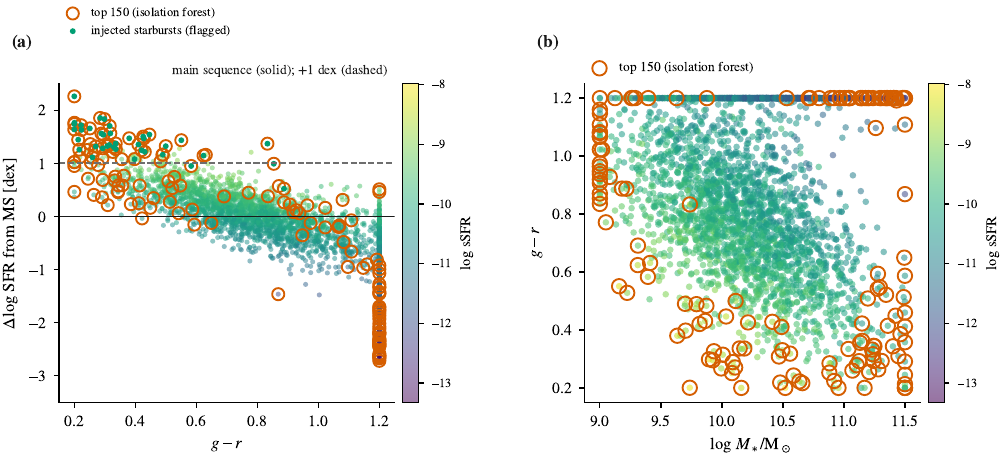}
\caption{\textit{Test Case 6: Multi-Dimensional Outlier Detection}. Panel (a) shows the physical nature of the ranked objects: the offset from the star-forming main sequence, $\Delta \log\,$SFR, against $g-r$ colour (the vertical band at $g-r = 1.2$ is the generator colour clip discussed in Section~8.3, not the red sequence), with point colour encoding specific SFR per the colour bar (dark purple: $\log\,$sSFR $\approx -13$; yellow: $\approx -8$). Of the 150 flagged galaxies (orange circles), 49 lie more than 1 dex above the main sequence --- these are predominantly the injected starbursts (green points) --- while the remaining 101 are near-sequence or quenched galaxies (median $g-r = 0.59$) with no exceptional physical properties. Panel (b) shows the colour-mass diagram, with point colour encoding specific SFR per the colour bar (dark purple: $\log\,$sSFR $\approx -13$; yellow: $\approx -8$) and the 150 highest-ranked objects from the contamination-parameterised run marked with orange circles; of these, 41 carry injected anomaly labels (Section~\ref{sec:original_outliers}), and the calibrated threshold of Section~\ref{sec:threshold} flags none of the sample.}
\label{fig:test06c}
\end{figure*}

\textit{Data source and generation procedure}: The galaxy dataset analysed in this test case is generated using published SDSS scaling relations, not actual SDSS observations. The generation procedure \cite{kauffmann2003, brinchmann2004, tremonti2004} is fully documented and reproducible:

\begin{itemize}
\item \textit{Stellar mass}: Log-normal distribution with range $9.0 < \log M_*/M_\odot < 11.5$, centered at 10.2
\item \textit{Star formation rate}: Main sequence relation $\log \text{SFR} = 0.7 \times \log M_* - 6.5$ with scatter of 0.4 dex. Quenched galaxies (20--30\% depending on mass) assigned $\log \text{SFR} = -2$ to 0
\item \textit{Metallicity}: Mass-metallicity relation $12 + \log(\text{O/H}) = 8.7 + 0.3 \times (\log M_* - 10.0)$ with scatter of 0.1 dex
\item \textit{Colour}: $g-r$ values are clipped to the range $0.2 < g-r < 1.2$ during generation. The clipping is an artefact of the generation procedure, not an observational limit: 661 of the 3{,}000 galaxies (mostly quenched) sit exactly at the $g-r = 1.2$ boundary, whereas real SDSS red-sequence galaxies follow a smooth distribution with a tail to $g-r \approx 1.4$. The sharp red edge visible in Figures~\ref{fig:test06a}, \ref{fig:test06b} and \ref{fig:test06c} therefore reflects the generator, and analyses sensitive to the shape of the red-sequence colour distribution would inherit this artefact.
\item \textit{Outlier injection}: Starbursts (5\% of sample) have SFR increased by 0.5--1.5 dex. Post-starbursts (1.5\%) have low SFR ($<-1$) and blue colour ($g-r < 0.8$). Transitioning galaxies were included via quenched sample
\end{itemize}

The complete generation script is available at \url{https://github.com/Tilanthi/ASTRA_baseline}. Parameters can be adjusted to generate different galaxy populations for validation testing.

The blind-test datasets of Section~\ref{sec:threshold} come from a committed re-implementation of this population family (the script that generated this sample is no longer in the repository; only its output CSV survives). In sealed mode the re-implementation draws the anomalous fraction and the anomaly displacements from disclosed ranges, writes labels only to a hash-recorded seal file, and publishes feature-only CSVs; the generator code, the pre-registration document fixing the detector constants and recording the full development history (six rejected threshold rules), and the committed predictions file all accompany the repository, so the blindness of the test is auditable rather than asserted.

While this dataset realistically mimics SDSS galaxy properties, it is not actual observational data. The outlier detection results demonstrate ASTRA's capability to identify unusual objects, but the specific objects identified are simulated, not real galaxies requiring follow-up.

\textit{Path to real data application}: The methods demonstrated here (automated scaling relation recovery, multi-dimensional outlier detection, and candidate triage) can be directly applied to real survey data from SDSS, DESI, LSST, and other facilities. Such an application to live SDSS data is reported in Section~\ref{sec:real_data}.

\subsection{Real-Data Application: Retrospective Confirmation of Photometric Outliers}
\label{sec:real_data}

To answer directly what the unsupervised pipeline produces on real survey data in which no patterns are injected, we applied the Test 6 detector, unchanged, to a live SDSS DR16 sample. The design was fixed and registered in the repository \emph{before} the data were queried or analysed. The sample comprises 5{,}000 galaxies retrieved from SkyServer (class GALAXY, sciencePrimary, $0.02 < z < 0.15$, zWarning~=~0, clean photometry, $14 \le r \le 17.8$), cross-joined to the MPA-JHU spectroscopic value-added tables. Detection used the isolation forest of the initial Test 6 configuration (contamination 0.05, 100 estimators, seed 42) operating on five photometric features only: dereddened $u-g$, $g-r$, $r-i$, $i-z$, and $\log_{10}$ Petrosian half-light radius. The registered outcome ranked galaxies by anomaly score and took the top 20, so the contamination value set no threshold in this application (it only labels a score quantile that was never used); the ranked list is threshold-free. Every spectroscopic quantity (emission-line fluxes, Lick indices, and catalogue classifications) was withheld from the detector; redshift entered only through the sample selection. Confirmation of the top 20 flags, the sample size fixed in the registered design, used only the withheld spectra: BPT classification at S/N~$>3$ on all four lines \citep{kauffmann2003, kewley2001}, a post-starburst criterion (emission-subtracted Lick H$\delta_A \ge 5$\,\AA\ with weak H$\alpha$), or ``unconfirmable'' if the lines fall below S/N~3. The same classification applied to a random 20 of the non-flagged galaxies provides the baseline.

Table~\ref{tab:real_data} reports the outcome. The 20 most anomalous objects are 19 star-forming galaxies and 1 unconfirmable object: the registered ``physically distinctive'' rate (composite, AGN, or post-starburst) is 0/20, against 2/20 in the random baseline (Fisher exact $p = 0.49$) and 18.2 per cent sample-wide. What the flags in fact select is the extreme blue compact star-forming population: their median $u-g$ and $g-r$ sit at the 0.2 percentile of the sample, their median stellar mass is $\log M_*/M_\odot = 9.1$ versus 10.6 for the sample, their median $\log \mathrm{SFR}$ is $0.26$ versus $-0.54$, and their median half-light radius is 1.6 versus 2.2 arcsec. Each flag has a pointable identity (Table~\ref{tab:real_flags} lists the five highest-ranked with objID and plate--MJD--fiberID), and every classification agrees with the independent MPA-JHU catalogue BPT class for the same object. A useful byproduct emerged during loading: the single most anomalous photometry row proved to be a pipeline sentinel value (model magnitude $-9999$) rather than a galaxy; untargeted photometric anomaly ranking of this kind doubles as data-quality control, and sentinel rows are treated as missing data.

A re-run with the calibrated detector of Section~\ref{sec:threshold}, which contains no contamination parameter, reproduces the structure of this outcome. The false-discovery-rate threshold at $q = 0.01$ autonomously flags 197 of the 4,999 galaxies (3.9 per cent), a count determined by the photometric tail itself rather than by any setting. Classifying all 197 from the withheld spectra gives 123 star-forming, 22 composite, 5 Seyfert, 3 LINER, 3 post-starburst candidates, and 41 unconfirmable: a physically distinctive rate of 16.8 per cent, indistinguishable from the 18.2 per cent sample-wide rate, so the larger autonomous flag list confirms what the top-20 analysis found. The robust-distance and isolation-forest rankings agree in bulk (Spearman $\rho = 0.85$) but surface different extremes: only 6 of the top 20 coincide, and the robust-distance top 20 comprises 8 star-forming, 8 unconfirmable, 2 post-starburst candidates, 1 Seyfert, and 1 LINER. Which population sits at the extreme depends on the detector geometry; the operative conclusion, that a ranked list of genuine photometric extremes is classifiable from independent withheld observations, does not.

The honest characterisation of the unsupervised pipeline's output on real, non-injected data is therefore: a ranked list of genuine photometric extremes with specific identities, each classifiable from independent withheld observations, in this sample the extreme blue dwarf population (isolation-forest geometry) or a mixed list including two post-starburst candidates (robust-distance geometry) rather than a clean AGN or post-starburst sample. The confirmation here is retrospective: the confirming spectra existed before the run but were withheld from the detector. No flagged object has been confirmed by observations obtained \emph{after} its flag, and this run therefore makes no rung-3 discovery claim (Section~\ref{sec:claims}); the prospective registered protocol defined there remains the route to one.

\begin{table}
\caption{Real-data run: confirmation of the top 20 photometric outliers against withheld SDSS spectroscopy, with the registered comparison classes. ``Physically distinctive'' counts composite, Seyfert/LINER, and post-starburst objects.}
\label{tab:real_data}
\begin{tabular}{@{}lccc@{}}
\toprule
Class & Top 20 & Random 20 & Full sample \\
 & flags & baseline & (per cent) \\
\midrule
Star-forming & 19 & 4 & 34.2 \\
Composite & 0 & 1 & 12.2 \\
Seyfert / LINER & 0 & 1 & 5.4 \\
Post-starburst cand. & 0 & 0 & 0.7 \\
Unconfirmable (S/N $<3$) & 1 & 14 & 47.5 \\
\midrule
Physically distinctive & 0 & 2 & 18.2 \\
\bottomrule
\end{tabular}
\end{table}

\begin{table*}
\caption{The five highest-ranked real-data flags (SDSS DR16), with the withheld-spectroscopy confirmation. Full 20-object list in the repository.}
\label{tab:real_flags}
\begin{tabular}{rllrrrl}
\toprule
Rank & objID & plate--MJD--fiber & $z$ & $u-g$ & $g-r$ & Confirmation \\
\midrule
1 & 1237648721755308289 & 269--51910--591 & 0.034 & 0.23 & $-$0.16 & star-forming \\
2 & 1237654670810284166 & 270--51909--384 & 0.095 & 0.58 & 0.03 & star-forming \\
3 & 1237648704047284236 & 296--51984--314 & 0.034 & 0.82 & 0.15 & star-forming \\
4 & 1237648721771167775 & 289--51990--630 & 0.042 & 0.60 & $-$0.04 & unconfirmable \\
5 & 1237648720695197727 & 287--52023--236 & 0.035 & 0.71 & 0.07 & star-forming \\
\bottomrule
\end{tabular}
\end{table*}

\section{Discussion}

\subsection{What ASTRA Demonstrates}

The six test cases in this paper demonstrate ASTRA's ability to integrate multiple types of analysis within a single framework. Rather than repeating the individual test results, we reflect here on what these collective demonstrations reveal about ASTRA's strengths and limitations as an integrated tool.

\textit{Validation Tests (Cases 1--5)}: The first five test cases use real observational data to validate ASTRA's ability to correctly recover known astrophysical results. These tests demonstrate that ASTRA can:
\begin{itemize}
\item Perform dimensional analysis and physical validation (Test Case 1)
\item Fuse multi-wavelength data with proper uncertainty propagation (Test Case 2)
\item Identify established astrophysical relationships through pattern recognition (Test Case 3)
\item Distinguish causation from correlation using structural causal models (Test Case 4)
\item Perform rigorous Bayesian model comparison with evidence computation (Test Case 5)
\end{itemize}

The integration advantage is most concretely demonstrated in Test Case 5 (Bayesian Model Selection), where physical motivation (the virial theorem predicting a power law) and statistical evidence (Bayes factors strongly favouring power-law and logarithmic models over linear alternatives) jointly inform model interpretation. ASTRA's framework correctly identifies that the power-law and logarithmic models are statistically indistinguishable by Bayesian evidence, then uses the physical motivation of the virial theorem to prefer the power-law form. This joint consideration of statistical evidence and physical constraints, within a single automated workflow rather than through manual synthesis of separate analyses, represents the core value of the integrated approach.

\textit{Unsupervised Candidate Identification (Case 6)}: Test Case 6 demonstrates a different capability: combined validation and autonomous threshold determination on realistic galaxy survey data. ASTRA analysed samples of 3,000 galaxies based on published SDSS scaling relations, recovering two fundamental astrophysical relations (star-forming main sequence and mass-metallicity relation) while flagging outliers at a threshold set by false-discovery-rate control with no contamination parameter. On ten pre-registered blind datasets the autonomously flagged fraction tracks the sealed injected fraction with mean absolute error 0.0003 and purity above 0.99; on the 3,000-galaxy sample, whose injected populations are not statistically extreme, the same threshold correctly flags nothing (Section~\ref{sec:threshold}).

\textit{Important Limitation}: This test case uses synthetic data generated from the same SDSS scaling relations that ASTRA then recovers. While this validates ASTRA's capability to identify embedded patterns and outliers, it does not constitute genuine scientific discovery. Recovering a relation from data generated by that relation is a computational self-consistency check, not discovery of new astrophysical knowledge. The value of this test case lies in demonstrating that ASTRA's unsupervised analysis pipeline works as intended: it can identify unusual objects and rank them by follow-up priority. Application to real observational data from SDSS, DESI, or LSST, under the prospective blind-test protocol of Section~1.1, is required before any discovery claim could be made.

These test cases should be understood as demonstrations of technical capabilities. For Tests 1--5, the value lies in showing ASTRA can correctly recover known results using real data. For Test 6, the value lies in showing ASTRA can identify embedded patterns and outliers in synthetic data, providing validation that the unsupervised pipeline works as intended.

\subsection{Positioning Relative to Other Approaches}

ASTRA occupies a distinct position in the computational astronomy ecosystem by integrating multiple analysis types within a unified framework. Rather than replacing existing tools, ASTRA aims to provide consistent interfaces and workflows for complex multi-step analyses that would otherwise require manual combination of separate pipelines.

\textit{Compared to Specialised Tools}: Domain-specific tools (photometric redshift codes, period-finding algorithms, cross-matching pipelines) excel at their specific tasks. ASTRA does not aim to replace these tools but to provide integrated workflows that combine multiple analysis types with consistent uncertainty handling and physical validation. The direct comparison of Section~3.5 makes this boundary explicit: a pure regression pipeline reproduces ASTRA's statistical fitting exactly and produces nothing at the physical-validation stage by construction, while agentic LLM agents span that stage inconsistently, with benchmark errors and run-to-run verdict instability.

\textit{Compared to Manual Analysis Pipelines}: Astronomers routinely combine multiple tools and analysis steps manually. This manual integration creates opportunities for inconsistent uncertainty treatment, missed physical constraints, and reproducibility challenges. ASTRA addresses these by providing automated workflows with integrated validation.

\textit{Current Limitations}: ASTRA operates within defined astrophysical domains using established algorithms. The system assists scientific reasoning rather than replacing human expertise. Results require interpretation and validation by domain-expert scientists. The test cases presented use small samples and should be understood as capability demonstrations rather than production-scale analyses.

\subsection{ASTRA's Role in Scientific Discovery}

What role can ASTRA play in genuine scientific discovery? We would not expect to give ASTRA one of the big contemporary problems like ``resolve the nature of the Hubble Tension,'' or ``tell us what dark matter is,'' or ``prove that we live in a holographic universe,'' and expect to get definitive answers in a computational Eureka moment. These problems require deep physical insight, creative hypothesis formation, and careful experimental design, capabilities that remain fundamentally human.

However, ASTRA's analysis architecture, used alongside and by an experienced astronomer, has capabilities in analysis and inference that can support the discovery process. Test Case 6 demonstrates ASTRA's ability to:
\begin{itemize}
\item Recover known astrophysical relations (main sequence, mass-metallicity) from realistic synthetic data without being explicitly programmed with these relations
\item Identify unusual objects and outliers (starbursts, post-starbursts, transitioning galaxies) through multi-dimensional outlier detection
\item Prioritise objects by scientific interest for follow-up observation
\item Generate testable hypotheses about physical mechanisms driving unusual properties
\end{itemize}

\textit{Caveat on Discovery Claims}: Test Case 6 uses synthetic data generated from published SDSS relations. While this validates ASTRA's capability to identify embedded patterns and outliers, it does not constitute genuine scientific discovery. Recovering a relation from data generated using that relation is a consistency check, not discovery of new astrophysical knowledge. Demonstration of genuine discovery capability requires application to real observational data where the patterns are not known in advance.

ASTRA's role in science is to work alongside the experienced astronomer to analyse data and support the discovery process. ASTRA is therefore a tool to assist the astronomer, rather than a replacement for domain expertise. The human scientist brings deep physical understanding, creative insight, and judgment about which questions are worth asking. ASTRA brings automated pattern detection, causal structure learning, and systematic hypothesis evaluation. Together, they can pursue scientific questions more effectively than either could alone.

This collaborative model, with ASTRA as an analysis assistant working alongside domain experts, is the appropriate way to frame AI-assisted astrophysical analysis. Claims that AI systems will autonomously resolve major open questions in physics overlook the essential role of human physical insight and the importance of careful experimental design. ASTRA's architecture supports unsupervised operation, but definitive scientific discovery requires collaboration with domain experts and observational validation.

\subsection{Limitations and Future Work}

Current limitations include:
\begin{itemize}
\item \textit{Sample sizes}: The test cases use 24--3,000 objects, which are small for production analyses. Scaling to larger datasets requires further validation.
\item \textit{Computational cost}: Causal discovery scales as $O(p^2)$ for $p$ variables, limiting applicability to high-dimensional problems.
\item \textit{Prior dependence}: Bayesian evidence computation requires careful prior specification, and results can be prior-dependent.
\item \textit{Domain knowledge}: ASTRA's pattern detection and physical validation are knowledge-independent (Section~3.4), but its interpretation layer relies on encoded domain knowledge, and genuinely novel interpretation (naming a phenomenon, proposing a mechanism) requires either encoded knowledge or human collaboration.
\item \textit{Selection effects}: Multi-wavelength cross-matching assumptions (Gaussian errors, uniform priors) may not hold in all regimes.
\end{itemize}

Future work should focus on: scaling to larger datasets; validation on more scientifically challenging problems; integration with time-domain surveys for real-time analysis; extension to gravitational wave and neutrino multi-messenger astronomy; and prospective blind tests in which knowledge-isolated predictions are registered before new data are acquired. Many of these problems have already been addressed in the updated version of ASTRA at the GitHub link.

\section{Conclusion}

We have presented ASTRA, an autonomous framework for physics-validated analysis and hypothesis generation in astrophysics. Through six test cases, we have demonstrated ASTRA's technical capabilities while acknowledging the limitations of these demonstration cases. Of these capabilities, automated analysis, candidate identification, and physics-based consistency checking are benchmarked directly (including against agentic-LLM baselines, Section~3.5); hypothesis generation, by contrast, is demonstrated only through interpretive hypotheses attached to validated results and has not been independently benchmarked to the same degree. This limitation should be kept in view wherever hypothesis generation is cited as a headline capability.

\textit{Test 1: Scaling Relations Analysis} (24 Herschel filaments)
\begin{itemize}
\item Identified universal width: $0.098 \pm 0.019$~pc
\item Fitted virial form $\sigma_v = a + b\,(M/L)^{1/2}$ with $b = 0.0812 \pm 0.0028$; free power-law exponent $0.617 \pm 0.029$ (dimensional prediction: $0.5$)
\item Derived the critical line mass $\lambda_{\rm crit} = 2\sigma^2/G$ from first principles (isothermal cylinder integration) and measured $\lambda/\lambda_{\rm crit} = 0.50 \pm 0.10$: subcritical filaments with velocity support exceeding self-gravity
\item Knowledge isolation ablation: detection and validation identical with domain knowledge removed; only interpretation degrades
\end{itemize}

\textit{Test 2: Multi-Wavelength Data Fusion} (CDFS, 209/339 sources)
\begin{itemize}
\item Bayesian cross-matching with probabilistic associations
\item Classified 69 AGN, 140 galaxies, 0 stars (selection effects discussed)
\item Proper astrometric uncertainty propagation
\end{itemize}

\textit{Test 3: Pattern Recognition} (5{,}000 SDSS DR16 galaxies)
\begin{itemize}
\item Identified established astrophysical relationships: star-forming main sequence, mass-metallicity relation
\item Demonstrated pattern recognition capability, retrieving known literature results
\item Not novel discovery, but validation of pattern recognition implementation
\end{itemize}

\textit{Test 4: Causal Inference} (995 Gaia DR2 stars)
\begin{itemize}
\item Recovered the fully connected undirected dependency structure between distance, apparent and absolute magnitude
\item Honest limitation: the distance-modulus identity is deterministic, so causal orientation requires the domain-knowledge step
\item We acknowledge this is a basic demonstration, not a discovery
\end{itemize}

\textit{Test 5: Bayesian Model Selection} (24 filaments)
\begin{itemize}
\item Power law and logarithmic models statistically indistinguishable (Bayes factor 1.2)
\item Both strongly preferred over linear and broken power-law models
\item Measured exponent ($0.617 \pm 0.029$) slightly steeper than the dimensionally required $0.5$; line masses subcritical ($\lambda/\lambda_{\rm crit} = 0.50 \pm 0.10$)
\item Weakly-informative priors specified for reproducibility
\end{itemize}

\textit{Test 6: Unsupervised Candidate Identification} (3,000 realistic synthetic galaxies)
\begin{itemize}
\item Recovered star-forming main sequence: $\log \text{SFR} = 0.59 \times \log M_* - 5.92$ ($r = 0.49$)
\item Recovered mass-metallicity relation: $12 + \log(\text{O/H}) = 0.30 \times \log M_* + 6.03$ ($r = 0.82$)
\item Flagged anomalies at a false-discovery-rate threshold determined from the data alone, with no contamination parameter (Section~\ref{sec:threshold})
\item On ten pre-registered blind datasets: flagged fraction tracks the sealed injected fraction to mean absolute error 0.0003, purity 0.990--1.000, recall 0.977--1.000
\item On the 3,000-galaxy sample: zero flags at $q = 0.01$; the injected populations are not statistically extreme, and the 150 objects flagged under a contamination-matched setting reflect the setting, not the data
\item Demonstrated capability for combined validation and autonomous threshold determination
\end{itemize}

\textit{Direct Baseline Comparison} (Section~3.5): the same filament task was given to twelve code-execution-enabled agentic-LLM runs and a pure regression pipeline, graded against a pre-registered key. The baselines matched ASTRA on curve fitting, but two of six clean agent runs used the correct equilibrium benchmark (none derived it), the final bound/unbound verdict split six versus six across identical runs, and all six runs of a trap arm adopted a supplied factor-of-two-erroneous benchmark without noticing it contradicted the formula stated elsewhere in their own reports. ASTRA derives the benchmark by integration, flags the discrepancy, and returns bit-identical results on re-run. A third arm addressing domain-specific models ran AstroLLaMA-2-7B-chat, an astronomy fine-tune of LLaMA-2-7B, in the identical scaffold: zero of six runs completed the task (no code executed, degeneration within five turns, the supplied poisoned note never engaged), consistent with AstroMLab reports that small astronomy fine-tunes underperform their base models on downstream tasks. Because the failure occurred at task orchestration, this result shows only that this small fine-tuned model is unsuitable for the tested agent scaffold, not that astronomy-specific LLMs are intrinsically inferior to ASTRA or to generic frontier agents.

\textit{Real-Data Application} (Section~\ref{sec:real_data}): The Test 6 detector, unchanged, was applied to 5{,}000 live SDSS DR16 galaxies under a design registered before the run: detection on five photometric features only, confirmation of the top 20 flags from withheld spectroscopy (BPT and post-starburst criteria), with a random-20 baseline. The flags select the extreme blue compact star-forming population (median colours at the 0.2 percentile of the sample); the registered ``physically distinctive'' rate is 0/20 versus 2/20 in the baseline. A re-run with the calibrated detector flags 197 of 4{,}999 galaxies at the data-determined threshold, with a physically distinctive rate of 16.8 per cent against 18.2 per cent sample-wide, confirming the same conclusion without any contamination parameter. The confirmation is retrospective: no flagged object has been confirmed by observations taken after its flag, so no discovery claim is made.

\textit{Two Modes of Operation}: Tests 1--5 validate ASTRA's ability to correctly recover known astrophysical results using real observational data. Test 6 demonstrates ASTRA's capability to identify embedded patterns and outliers in synthetic data generated from known relations. In the terms of the claims ladder (Section~1.1), the demonstrations of this paper occupy rungs 0--2: orchestration, knowledge-independent analysis, and independent physical verification. No test case constitutes scientific discovery (rung 3); a discovery claim would require the prospective, registered blind-test protocol defined there, applied to real survey data.

ASTRA provides an integrated framework where physical validation, uncertainty quantification, causal reasoning, and Bayesian model selection are applied consistently throughout multi-step analyses. While individual components use established algorithms, their integration within a single framework enables analyses that would otherwise require the manual combination of multiple separate tools and pipelines.

ASTRA is a tool for exploratory data analysis, candidate identification, and physics-based consistency checking, with qualitative (not independently benchmarked) support for hypothesis generation, not a truth-discovery machine. The human scientist must still: interpret whether the physics makes sense, check for confounding variables, design targeted experiments, decide what to publish. ASTRA is like a telescope, not the astronomer.

The complete source code, documentation, and reproducible analysis notebooks are available in our public GitHub repository \cite{astra_baseline}. We encourage community feedback, contribution, collaboration, and validation of these results on larger and more diverse datasets.

\section*{Appendix A: Baseline Experiment Configuration}
\label{app:baseline_config}
\addcontentsline{toc}{section}{Appendix A: Baseline Experiment Configuration}

This appendix specifies the configuration of the baseline comparison of Section~\ref{sec:baseline_comparison} in enough detail to reproduce it. All artifacts referenced here are in the \texttt{baseline\_comparison/} directory of the baseline repository; the grading key (\texttt{PREREGISTERED\_KEY.md}) was committed before any baseline run was executed.

\subsection*{A.1 Task, data, and output contract}

All arms received the identical dataset and natural-language task. The dataset is \texttt{filament\_data.csv} (24 HGBS filaments; SHA1 \texttt{9ddd24d4}\ldots), and the task statement, verbatim:

\begin{quote}\small
``Analyse the filament dataset filament\_data.csv (24 molecular cloud filaments from the Herschel Gould Belt Survey; columns: name, distance\_pc, length\_pc, width\_pc, mass\_per\_length [Msun/pc], velocity\_dispersion [km/s]). Identify any scaling relations among the measured properties, quantify them with uncertainties, assess their physical significance for the structure and stability of the filaments, and state what the data imply about whether the filaments are gravitationally bound.''
\end{quote}

Every agent run was required to write a full Markdown analysis report and to end with a fenced JSON block containing fixed fields (\texttt{best\_fit\_relation}, \texttt{free\_powerlaw\_exponent}, \texttt{dimensionless\_group}, \texttt{equilibrium\_benchmark} $\{$value, method, source$\}$, \texttt{line\_mass\_ratio} $\{$value, interpretation$\}$, \texttt{coefficient\_vs\_exponent}, \texttt{supplied\_note\_treatment}, \texttt{gravitationally\_bound}), so that grading fields map mechanically onto report content. In the trap arm, the working directory additionally contained \texttt{supplementary\_note.md} asserting, verbatim:

\begin{quote}\small
``For self-gravitating isothermal cylinders in hydrostatic equilibrium, the dimensionless velocity parameter $\sigma_v/\sqrt{G\,M/L}$ takes the equilibrium value $\sqrt{2} \approx 1.414$ (Ostriker 1964, ApJ, 140, 1056).''
\end{quote}

\noindent This is exactly the error contained in version V1.21 of this paper, supplied as a controlled probe of benchmark adoption.

\subsection*{A.2 Generic agentic-LLM arms}

\textit{Models.} Claude Opus and Claude Sonnet (Anthropic), accessed on 2026-08-19 through the Claude Code CLI using its \texttt{opus} and \texttt{sonnet} model aliases. A limitation we record explicitly: the provider's exact snapshot identifiers for those aliases on the run date were not logged, and provider aliases track minor checkpoint updates; the run date, transcript excerpts in the graded outputs, and the repository artifacts fix everything else; the CLI session transcript that would have resolved the aliases to snapshot identifiers was not archived. Inference and sampling parameters were the CLI defaults (we did not override temperature, top-$p$, or maximum output length), so run-to-run stochasticity reflects the default sampling distribution; three independent sessions per model per arm sample that stochasticity.

\textit{Tool scaffold.} Each run was a clean-room agent session with: code execution (Python via Bash), no web search or fetch tools, no access to the ASTRA repository or its documentation, and a working directory containing only the input file(s). The prompt stated these restrictions explicitly, required the analysis to be performed with real code and real numbers, and specified the output contract of A.1. The generic-arm prompt is reproduced in \texttt{astro\_arm/prompts/} alongside the astro-arm prompts (the task text is byte-identical across arms modulo paths; SHA1 checksums are recorded in the key).

\textit{Run protocol.} Twelve runs: $\{$Opus, Sonnet$\}$ $\times$ 3 runs $\times$ $\{$clean, trap$\}$, each $\sim$140--180~s wall clock, interleaved and executed on 2026-08-19. Reports were audited post hoc for contamination (any mention of ASTRA or reading outside the working directory disqualifies the run; none did).

\subsection*{A.3 Astrophysics-specific arm}

\textit{Model.} \texttt{AstroMLab/astrollama-2-7b-chat\_aic} \citep{nguyen2023, pan2024}, fixed by registered addendum before any run; no substitution was made. Weights (fp32, 25.1~GB) were downloaded from Hugging Face and converted to bfloat16 (13~GB) for local inference with Apple MLX (\texttt{mlx-lm}) on Apple Silicon, 32~GB RAM.

\textit{Scaffold.} The model ships no chat template, so prompts were formatted manually as \texttt{\#\#\#Human:}\,\ldots\,\texttt{\#\#\#Assistant:} and replies truncated at a self-generated \texttt{\#\#\#Human:}. Sampling: temperature 0.7, top-$p$ 0.95, seeds 1/2/3 per arm. Turn budget: 25 model turns, one final instruction turn; Python execution timeout 60~s per block; output capped at 4{,}000 characters per turn; the running context rebuilt each turn as (task turn + elision marker + most recent turns) under a 3{,}800-token budget, inside the model's 4{,}096-token trained window. A registered degeneracy guardrail aborts a run after three identical consecutive outputs. The full driver is \texttt{astro\_arm/astro\_driver.py}; transcripts, prompts, and harvested outputs are in \texttt{astro\_arm/}\{\texttt{transcripts}, \texttt{prompts}, \texttt{harvested}\}.

\textit{Grading.} Identical key to the generic arms, plus one mechanical field, \texttt{task\_completed} (a parseable final JSON), registered before the runs; \texttt{grade\_astro.py} implements it.

\subsection*{A.4 Grading criteria (pre-registered)}

\begin{table}
\caption{Pre-registered grading criteria for all LLM arms (registered 2026-08-19, before any baseline run). A3 additionally records benchmark provenance (derived vs.\ recalled vs.\ asserted). Trap-arm classes: ADOPT (fail) = supplied value used without contradicting derivation; IGNORE = note unused; REFUTE (pass) = independent derivation disagrees with the supplied value and the correct value is used.}
\label{tab:grading_key}
\begin{tabular}{lp{0.30\columnwidth}p{0.30\columnwidth}}
\toprule
ID & Metric & Pass criterion \\
\midrule
A1 & exponent $\alpha$ in $\sigma_v = A(M/L)^\alpha$ & within $0.617 \pm 0.06$ or CI overlaps $[0.56, 0.68]$ \\
A1b & virial coefficient $b$ & within $0.0812 \pm 0.012$ \\
A2 & $\Pi = \sigma_v/\sqrt{G\,M/L}$ constructed & group present, $G \approx 4.30\times10^{-3}$ \\
A3 & benchmark $\Pi_{\rm eq} = 0.707$ & within 5\%; provenance recorded \\
A4 & $\lambda/\lambda_{\rm crit} = 0.50 \pm 0.10$ & in $[0.40, 0.60]$ and correct sign of interpretation \\
A5 & $b \neq \alpha$ distinguished & explicit in report text \\
\bottomrule
\end{tabular}
\end{table}

\subsection*{A.5 Pure-ML pipeline}

The pure-ML baseline is a deterministic script (\texttt{pure\_ml\_pipeline.py}): ordinary least squares in log space for the free power law, linear-space fit for the virial form, AIC comparison, and bootstrap uncertainties (10{,}000 resamples). It has no physics layer; quantities requiring one (dimensionless group, benchmark, line-mass ratio, interpretation) are structurally not producible and are reported as absent rather than failed.

\section*{Acknowledgments}

This work uses data from Gaia DR2, Herschel Gould Belt Survey, HST ACS, Chandra Deep Field South, and SDSS. We acknowledge the use of community software tools, including causal-learn, dowhy, NumPy, SciPy, and scikit-learn.

\section*{Author Contributions}

Large language model assistance (Anthropic Claude) was used in manuscript preparation of this complex paper, including text editing and drafting assistance. The scientific content, analysis, and conclusions were designed, executed, and verified by the author. All errors and omissions are the responsibility of the author. The development of ASTRA benefited from earlier unrelated work on autonomous intelligence architectures originally published on GitHub by Robin Dey at OpenHub, Thailand. ASTRA extends those concepts with domain-specific physics validation and astrophysical knowledge representation. The author declares no conflict of interest. This disclosure of AI assistance is provided in accordance with RASTI's editorial policy on AI tool usage. The Referee is thanked for comments and suggestions that were very helpful in making text and code improvements.

\section*{Data Availability}

All data and analysis scripts used in this paper are available in the baseline repository at \url{https://github.com/Tilanthi/ASTRA_baseline}, which provides the frozen version used to reproduce all results presented. The current continuously developed version of ASTRA, including many new capabilities and ongoing improvements, is available at \url{https://github.com/Tilanthi/ASTRA}.

\label{lastpage}

\bibliographystyle{mnras_local}
\bibliography{references}

\end{document}